\pdfoutput=1
\documentclass[11pt]{article}
\usepackage[margin=1in]{geometry}
\usepackage[T1]{fontenc}
\usepackage{lmodern,microtype}
\usepackage{amsmath,amssymb,amsthm}
\usepackage{xcolor}
\usepackage{tikz}
\usetikzlibrary{arrows.meta,shapes.misc,patterns,calc}
\usepackage[colorlinks=true,linkcolor=blue!50!black,citecolor=blue!50!black,urlcolor=blue!50!black,bookmarksnumbered=true]{hyperref}
\usepackage{bookmark}
\hypersetup{pdftitle={The massive phase of the two-dimensional Heisenberg model is stable under small anisotropy},pdfauthor={Yan Ru Pei},pdfkeywords={classical Heisenberg model, O(n) model, anisotropy, exponential decay, Berezinskii-Kosterlitz-Thouless transition, Polyakov conjecture, correlation inequalities, disagreement percolation}}
\newtheorem{theorem}{Theorem}[section]
\newtheorem{proposition}[theorem]{Proposition}
\newtheorem{lemma}[theorem]{Lemma}
\newtheorem{corollary}[theorem]{Corollary}
\theoremstyle{definition}

\theoremstyle{remark}
\newtheorem{remark}[theorem]{Remark}
\numberwithin{equation}{section}
\newcommand{\Z}{\mathbb Z}
\newcommand{\R}{\mathbb R}
\newcommand{\HH}{\mathbb H}
\newcommand{\Pp}{\mathbb P}
\newcommand{\E}{\mathbb E}
\newcommand{\1}{\mathbf 1}
\newcommand{\lr}{\leftrightarrow}
\newcommand{\Cov}{\operatorname{Cov}}
\newcommand{\dist}{\operatorname{dist}}
\newcommand{\coll}{\operatorname{coll}}
\newcommand{\Jv}{\mathbf J}
\newcommand{\Sys}{\mathfrak S}
\newcommand{\an}{\mathrm{an}}
\newcommand{\iso}{\mathrm{iso}}
\makeatletter\g@addto@macro\bfseries{\boldmath}\makeatother

\title{The massive phase of the two-dimensional Heisenberg model\\ is stable under small anisotropy}
\author{Yan Ru Pei\\[0.3em]\small\href{mailto:yanrpei@gmail.com}{\texttt{yanrpei@gmail.com}}}
\date{November 2026}

\begin{document}
\maketitle

\begin{abstract}
Polyakov predicted that the classical Heisenberg model on $\mathbb{Z}^2$ has exponentially decaying
correlations at every temperature, and proofs of this have recently been posted. We show that the
massive phase survives small anisotropy. For every inverse temperature $\beta$ there is
$\delta(\beta)>0$ such that the $S^2$-valued model with couplings
$\sum_iJ^i\sigma_x^i\sigma_y^i$ has exponentially decaying spin two-point functions, uniformly in finite
subgraphs with free boundary conditions and in edge strengths up to $\beta$, whenever all
$|J^i-1|\le\delta(\beta)$. The same holds for aligned single-ion anisotropy and for nearly isotropic $O(n)$
models with $n\ge3$. Consequently, for the easy-plane model with couplings
$\sigma^x\sigma^x+\sigma^y\sigma^y+\lambda\sigma^z\sigma^z$, the onset of the
Berezinskii--Kosterlitz--Thouless phase, proved by Garban and de Montgolfier for every
$\lambda\in(-1,1)$, moves to zero temperature as $\lambda\uparrow1$. These results rest on the annulus
crossing bound of Ding, Huang and Yang, used on one finite annulus where the anisotropy is a bounded
perturbation, and on a sign-cluster argument with Ginibre's inequality for anisotropic planar energies.
Explicit constants of another isotropic proof give decay for $\beta\le c\,|\log(1-\lambda)|^{1/2}$. For
$n\ge4$ and $|J^i-1|\le\delta$ with $\delta$ small, at $\beta\le(1-\varepsilon)|\log\delta|/(2\pi)$ spin
one-point functions decay exponentially away from the boundary under every boundary condition, and
two-point functions decay, both on the scale $e^{\pi\beta+O(\sqrt{\beta\log\beta})}$. This is the $O(4)$
threshold: for $n=4$ it matches the renormalization-group crossover to leading order; for $n\ge5$ it lies
below the prediction. This part is conditional on estimates of a recent manuscript on the $O(4)$ model,
which the proof transfers to anisotropic couplings at one scale; only odd observables transfer.
\end{abstract}

\section{Introduction}\label{sec:intro}

\subsection{The model and the question}

Let $G=(V,E)$ be a finite subgraph of the nearest-neighbour graph of $\Z^2$, let $b\in[0,\infty)^E$
be edge strengths, and let $\Jv=(J^1,J^2,J^3)\in[0,\infty)^3$. The \emph{anisotropic Heisenberg
model} is the probability measure on $(S^2)^V$ with density
\begin{equation}\label{eq:model}
\frac1{Z}\exp\Bigl(\sum_{\{x,y\}\in E}b_{xy}\sum_{i=1}^3J^i\sigma_x^i\sigma_y^i\Bigr)
\end{equation}
with respect to the product of uniform probability measures on $S^2$; we write
$\langle\cdot\rangle_{G,b,\Jv}$ for its expectation, and $\langle\cdot\rangle_{\beta,\lambda}$ when $b\equiv\beta$ and
$\Jv=(1,1,\lambda)$ (the $\lambda$-model below). Except in Theorem~\ref{thm:phys} and Section~\ref{sec:phys},
all boundary conditions in the statements of results are free. For $\Jv=(1,1,1)$ this is the classical Heisenberg model,
and Polyakov~\cite{Polyakov1975} predicted exponential decay of correlations at every $\beta$. Proofs have
recently been posted~\cite{OpenAI215,DHY2026}. Write $\langle\cdot\rangle^{(n)}_{G,b}$ for the
$S^{n-1}$-valued model with density proportional to $\exp(\sum_{\{x,y\}\in E}b_{xy}\sigma_x\cdot\sigma_y)$. In the
form uniform in subgraphs and strengths, for every $n\ge3$ and $\beta<\infty$ there are $A,m>0$ with
\begin{equation}\label{eq:polyakov}
0\le\langle\sigma_x\cdot\sigma_y\rangle^{(n)}_{G,b}\le Ae^{-m\|x-y\|_2}
\end{equation}
for every finite subgraph $G$ of $\Z^2$ and every $b\in[0,\beta]^E$~\cite[Theorem~1.1]{OpenAI215}. Ding,
Huang and Yang prove exponential decay of the free-boundary two-point function in the boxes $B_L$ of side
length $L$ centred at the origin, for points of $B_{L/2}$, and of the one-point function under constant boundary conditions, for every
$n\ge3$~\cite[Theorem~1.1]{DHY2026}. Their annulus crossing bound~\cite[Theorem~5.2]{DHY2026}, combined
with the sign-cluster argument of Sections~\ref{sec:clusters}--\ref{sec:main}, also
gives~\eqref{eq:polyakov} in the uniform form (Remark~\ref{rem:uniform}).

The \emph{$\lambda$-model} has $b\equiv\beta$ and $\Jv=(1,1,\lambda)$. For $\lambda>1$ it has an Ising-type
ordered phase at low temperature~\cite{Malyshev1975,KunzPfisterVuillermot1976,FrohlichLieb1978}.
For $\lambda<1$, Bricmont, Lebowitz and Pfister conjectured a Berezinskii--Kosterlitz--Thouless (BKT)
phase at low temperature~\cite{BricmontLebowitzPfister1985}, which Dunlop proved for small
$\lambda$~\cite{Dunlop1985} and Garban and de Montgolfier proved for every
$\lambda\in(-1,1)$~\cite[Theorem~1.2]{GarbanMontgolfier2026}: there is $\bar\beta(\lambda)$ such
that for $\beta\ge\bar\beta(\lambda)$ the in-plane two-point function in free or Dirichlet
infinite-volume limits, in particular in free-boundary limits, decays at most polynomially. The conjectured phase diagram
\cite[Figure~2]{GarbanMontgolfier2026} has a curve $\lambda_c(T)$ separating the BKT phase from an
exponential phase and approaching $\lambda=1$ as $T\to0$, with renormalization-group prediction
$\lambda_c(T)\approx1-e^{-c/T}$~\cite{Khokhlachev1976,HikamiTsuneto1980,Pires1996}. For the
hemisphere model, in which the spins are restricted to the upper half of $S^2$, Garban and de
Montgolfier proved that $\bar\beta(\lambda)\to\infty$ as $\lambda\to1$, with exponential decay whenever
$\exp(C\beta)<1/(1-\lambda)$~\cite[Theorem~1.7]{GarbanMontgolfier2026}. For the full sphere, whether
$\bar\beta(\lambda)\to\infty$ as $\lambda\uparrow1$ has been open, and it is at least as hard as Polyakov's prediction: by a correlation inequality of
Dunlop~\cite{Dunlop1976} the in-plane correlation is nonincreasing in
$\lambda$~\cite[Proposition~A.1]{GarbanMontgolfier2026}, so exponential decay at some $\lambda<1$ gives it at
$\lambda=1$.

\subsection{Results near isotropy}

\begin{theorem}\label{thm:main}
For every $\beta\in(0,\infty)$ there are $\delta=\delta(\beta)>0$ and $A,m>0$ such that, for every finite
subgraph $G$ of $\Z^2$, every $b\in[0,\beta]^E$, every $\Jv\in[1-\delta,1+\delta]^3$, every
$i\in\{1,2,3\}$ and all $x,y\in V$,
\[
0\le\langle\sigma_x^i\sigma_y^i\rangle_{G,b,\Jv}\le Ae^{-m\|x-y\|_2}.
\]
One can take $A=A_0:=1/(4(1-2^{-1/25}))$, an absolute constant.
\end{theorem}

The off-diagonal functions $\langle\sigma_x^i\sigma_y^j\rangle$, $i\ne j$, vanish, because~\eqref{eq:model} is
invariant under the sign change of the $i$th coordinate at every site, so Theorem~\ref{thm:main} bounds every
spin two-point function. The bounds are uniform in the volume, so they pass to free-boundary limits
(Section~\ref{sec:cor}). Applied with $b\equiv\beta'\le\beta$, Theorem~\ref{thm:main} gives the following.

\begin{corollary}\label{cor:lambda}
If $|1-\lambda|\le\delta(\beta)$, the $\lambda$-model has exponentially decaying two-point functions at every
inverse temperature $\beta'\in(0,\beta]$, uniformly in finite free volumes and in free-boundary limits.
Consequently every threshold $\bar\beta(\lambda)$ as in~\cite[Theorem~1.2]{GarbanMontgolfier2026} satisfies
$\bar\beta(\lambda)\to\infty$ as $\lambda\uparrow1$; and if $0\le\lambda-1\le\delta(\beta)$, there is no long-range
order in free-boundary limits at any $\beta'\le\beta$.
\end{corollary}

The same argument applies to other small anisotropies for which the planar energies in the proof stay
nonnegative combinations of cosines (Section~\ref{sec:clusters}); for single-ion terms this is the
alignment condition below.

\begin{theorem}\label{thm:ext}
There is $\delta'(\beta)\in(0,\delta(\beta)]$, given in Section~\ref{sec:ext}, such that the following hold,
with the constants $A,m$ of Theorem~\ref{thm:main}, which do not depend on $n$.
\begin{enumerate}
\item[(a)] The conclusion of Theorem~\ref{thm:main} holds for the density
$\exp\bigl(\sum_eb_e\sum_iJ^i\sigma_x^i\sigma_y^i+\sum_x\sum_ic_{x,i}(\sigma_x^i)^2\bigr)$ with
$\Jv\in[1-\delta',1+\delta']^3$ and $c_{x,i}\in[-\beta\delta',\beta\delta']$, provided the anisotropies are
\emph{aligned}: some permutation $\pi$ of $\{1,2,3\}$ satisfies $J^{\pi1}\ge J^{\pi2}\ge J^{\pi3}$ and
$c_{x,\pi1}\ge c_{x,\pi2}\ge c_{x,\pi3}$ for every $x$.
\item[(b)] For every $n\ge3$, the conclusion holds for the $S^{n-1}$-valued model with density
proportional to $\exp\bigl(\sum_eb_e\sum_{i=1}^nJ^i\sigma_x^i\sigma_y^i\bigr)$ and
$\Jv\in[1-\delta,1+\delta]^n$, with $\delta=\delta(\beta)$ of Theorem~\ref{thm:main}.
\end{enumerate}
\end{theorem}

Part~(a) covers the confining potential $\tilde H_\varepsilon=-\sum\langle\sigma_i,\sigma_j\rangle+\varepsilon\sum(\sigma_i^z)^2$
of~\cite[Definition~1.8(4)]{GarbanMontgolfier2026} for $|\varepsilon|\le\delta'(\beta)$, for which Garban and de
Montgolfier state a BKT phase at large $\beta$ when $\varepsilon>0$, with an outline of
proof~\cite[Theorem~1.9]{GarbanMontgolfier2026}.
It also covers the $\lambda$-model with an easy-plane single-ion term when $\lambda\le1$, or an easy-axis one when
$\lambda\ge1$. Finally, with free boundary conditions the in-plane correlation
$\langle\sigma_x^1\sigma_y^1\rangle_{\beta,\lambda}$ is nonincreasing and the out-of-plane correlation
$\langle\sigma_x^3\sigma_y^3\rangle_{\beta,\lambda}$ nondecreasing in $\lambda\in[0,\infty)$
\cite[Proposition~A.1]{GarbanMontgolfier2026}. Combined with~\eqref{eq:polyakov}, this bounds the weakly
coupled components at every $\lambda$ and $\beta$.

\begin{corollary}\label{cor:mono}
For every $\beta>0$ and with free boundary conditions on finite boxes of $\Z^2$: if $\lambda\in[0,1]$
then $\langle\sigma_x^3\sigma_y^3\rangle_{\beta,\lambda}\le\frac A3e^{-m\|x-y\|_2}$, and if $\lambda\ge1$ then
$\langle\sigma_x^1\sigma_y^1\rangle_{\beta,\lambda}\le\frac A3e^{-m\|x-y\|_2}$, with $A,m$ from~\eqref{eq:polyakov} for
$n=3$; for instance $A=3A_0$ and $m=m_0(\beta)$ of Remark~\ref{rem:uniform}.
\end{corollary}

\subsection{Rates}\label{sec:rates}

How fast must $\delta(\beta)$ shrink as $\beta\to\infty$? In the renormalization-group picture the
anisotropy is negligible on short scales, where the model looks isotropic: its energy on a correlated
block of side $\ell$ is of order $\beta\delta\ell^2$. It takes over at the scale $\xi(\beta)$ of the isotropic
correlation length once $\beta\delta\xi(\beta)^2$ is of order one; the crossover is at
$\beta\delta\xi(\beta)^2\asymp1$. For $n=3$
the predicted correlation length is $\xi\approx e^{2\pi\beta}/\beta$, so the crossover is at
$\beta\approx(4\pi)^{-1}|\log\delta|$; for the $\lambda$-model this is the predicted BKT threshold
$\beta_{\mathrm{BKT}}(\lambda)\approx(4\pi)^{-1}\log(1/(1-\lambda))$~\cite{HikamiTsuneto1980,IrkhinKatanin1999}.
For $n=4$, \cite[Theorem~1.1]{OpenAIO4} gives $\xi\asymp e^{\pi\beta}/\sqrt\beta$ for large $\beta$, so the
crossover is at $\beta\approx(2\pi)^{-1}|\log\delta|$. For $n\ge4$ we reach this $O(4)$ threshold; for
$n=4$ it is the crossover, and for $n\ge5$ it lies below the $O(n)$ prediction.

\begin{theorem}\label{thm:phys}
Assume the estimates (I1)--(I3) of Section~\ref{sec:o4inputs}, taken from~\cite{OpenAIO4}: a stiffness bound
that fixes the coarse scale, a deterministic profile construction, and a bad-site estimate conditional on the
outside rings. There are absolute constants $A_*,C_\Xi<\infty$ such that, with
\[
\Xi(\beta):=\exp\Bigl(\pi\beta+C_\Xi\sqrt{(1+\beta)\log(2+\beta)}\Bigr),
\]
for every $\varepsilon\in(0,1)$ there is $\delta_0(\varepsilon)>0$ with the following property. Let
$\delta\in(0,\delta_0(\varepsilon))$, $0<\beta\le(1-\varepsilon)|\log\delta|/(2\pi)$ and $n\ge4$. Let $V\subset\Z^2$
be finite, $E$ a set of nearest-neighbour edges between sites of $V$, $P\subset V$ a set of pinned sites with
arbitrary values $\bar\sigma\in(S^{n-1})^P$, and, for each $e\in E$, $b_e\in[-\beta,\beta]$ and $m_e\in\R^n$ with
$|m_e^k-b_e|\le\beta\delta$ for all $k$; for instance $m_e=b_e\Jv$ with $b_e\in[0,\beta]$ and
$\Jv\in[1-\delta,1+\delta]^n$. Let $\langle\cdot\rangle$ denote expectation under the probability measure on
$(S^{n-1})^{V\setminus P}$ with density proportional to
$\exp\bigl(\sum_{\{x,y\}\in E}\sum_km^k_{xy}\sigma_x^k\sigma_y^k\bigr)$ with respect to the product of uniform
measures, where $\sigma_p=\bar\sigma_p$ for $p\in P$. Then, for all $i,j\in\{1,\dots,n\}$ and $x,y\in V\setminus P$,
\begin{align*}
\text{(i)}\quad&|\langle\sigma_x^i\rangle|\le A_*\exp\bigl(-\dist_\infty(x,P)/\Xi(\beta)\bigr),\\
\text{(ii)}\quad&|\langle\sigma_x^i\sigma_y^j\rangle|\le A_*\exp\bigl(-\min\{\|x-y\|_\infty,\dist_\infty(x,P)\}/\Xi(\beta)\bigr),
\end{align*}
with $\dist_\infty(x,\emptyset)=\infty$. With free boundary conditions ($P=\emptyset$) this gives
$|\langle\sigma_x^i\sigma_y^j\rangle|\le A_*e^{-\|x-y\|_\infty/\Xi(\beta)}$, and $\langle\sigma_x^i\sigma_y^j\rangle=0$ for $i\ne j$.
On square tori $(\Z/L\Z)^2$ with such couplings on all nearest-neighbour edges,
$|\langle\sigma_x^i\sigma_y^j\rangle|\le e\,A_*e^{-d_T(x,y)/\Xi(\beta)}$, with $d_T$ the cyclic $\ell^\infty$ distance. Every
infinite-volume Gibbs measure $\mu$ for such couplings on the edges of $\Z^2$ has $\mu(\sigma_x^i)=0$ and
$|\mu(\sigma_x^i\sigma_y^j)|\le A_*e^{-\|x-y\|_\infty/\Xi(\beta)}$.
\end{theorem}

For $n=4$, Theorem~\ref{thm:main4} gives the same bounds for every odd observable of one spin and every
anisotropy matrix $M_e$ with $\|M_e-b_eI\|_{\mathrm{op}}\le\beta\delta$. The condition on $\beta$ can be relaxed
to $\beta\le(2\pi)^{-1}(|\log\delta|-K\sqrt{|\log\delta|\log|\log\delta|})$ for small $\delta$
(Lemma~\ref{lem:threshold}). The proof uses the condition only in the form $\beta\delta\Xi(\beta)^2\le c_1$:
the crossover condition with $\xi$ replaced by $\Xi(\beta)$, which differs from
$e^{\pi\beta}/\sqrt\beta$ by a factor $e^{O(\sqrt{\beta\log\beta})}$. $\Xi(\beta)$ has the form of the length
scale of~\cite[(1.7), (3.1)]{OpenAIO4}, and $1/\Xi(\beta)$ that of the lower bound on the $O(4)$ mass
in~\cite[Corollary~1.2]{OpenAIO4}. So, to leading order, the threshold of
Theorem~\ref{thm:phys} is the crossover. It applies, for example, to the easy-axis model
$\Jv=(1+\delta,1,1,1)$, for which the renormalization-group picture predicts Ising-type order at larger
$\beta$: no Gibbs measure is magnetized at $\beta\le(1-\varepsilon)|\log\delta|/(2\pi)$. For $n\ge5$ the
threshold is that of $O(4)$, below the prediction $(n-2)|\log\delta|/(4\pi)$. Theorem~\ref{thm:phys} holds
under every boundary condition and for signed couplings, but only for odd observables: the analogous
one-point statement for even observables with vanishing $S^3$-average, such as $(\sigma_x^1)^2-\frac14$ for
$n=4$, fails already on two sites (Remark~\ref{rem:twosite}).

For $n=3$ we have a weaker rate, from the explicit constants of~\cite{OpenAI215}; the argument behind
Theorem~\ref{thm:phys} does not reach $n=3$ (Remark~\ref{rem:scope}).

\begin{theorem}\label{thm:rate}
There is an absolute constant $C<\infty$ such that $\delta(\beta)$ in Theorem~\ref{thm:main} can be
taken with $\delta(\beta)\ge{\exp(-C(1+\beta^2))}$. Hence there is an absolute $c>0$ such that, for
$\lambda$ close to $1$, the $\lambda$-model has exponentially decaying correlations at every
$\beta\le c\,{|\log|1-\lambda||}^{1/2}$.
\end{theorem}

The exponent $\frac12$ comes from the factor $\beta^2$ in the rotation estimate of
\cite[Proposition~2.2]{OpenAI215} (Section~\ref{sec:rate}); the prediction is of order $|\log|1-\lambda||$.

\subsection{Sources}\label{sec:sources}

Theorems~\ref{thm:main} and~\ref{thm:ext} and Corollaries~\ref{cor:lambda} and~\ref{cor:mono} rest
on~\cite[Theorem~5.2]{DHY2026} and the loop representation of~\cite[\S\S2--3]{DHY2026}, used through
Proposition~\ref{prop:dhy} with non-constant couplings (Remark~\ref{rem:dhy}); the sign-cluster arguments
of Campbell and Chayes~\cite{CampbellChayes1998} and of~\cite[\S4]{OpenAI215} are written out in
Sections~\ref{sec:clusters} and~\ref{sec:main}, and these results use no statement of~\cite{OpenAI215}.
Corollaries~\ref{cor:lambda} and~\ref{cor:mono} also use~\cite[Theorem~1.2 and
Proposition~A.1]{GarbanMontgolfier2026}. Theorem~\ref{thm:rate} uses~\cite[Proposition~3.1]{OpenAI215} in
place of Proposition~\ref{prop:dhy} (Section~\ref{sec:rate}) and the explicit constants in its proof,
namely~\cite[Lemma~2.1, Proposition~2.2, Lemma~3.2, (3.10), (3.16)--(3.20)]{OpenAI215}.
Theorem~\ref{thm:phys} uses~\cite[(3.54) and (3.58)]{OpenAIO4} and the construction
of~\cite[\S3.10.1]{OpenAIO4}, in the form stated in Section~\ref{sec:o4inputs}; in~\cite{OpenAIO4}, (3.58) is
proved from the stiffness bound in (3.54) and~\cite[(3.10), (3.11), (3.13), (3.14)]{OpenAIO4}, that is,
from the stiffness iteration of~\cite[\S\S3.1--3.8]{OpenAIO4}. None of~\cite{DHY2026}, \cite{OpenAI215}
and~\cite{OpenAIO4} has been refereed.

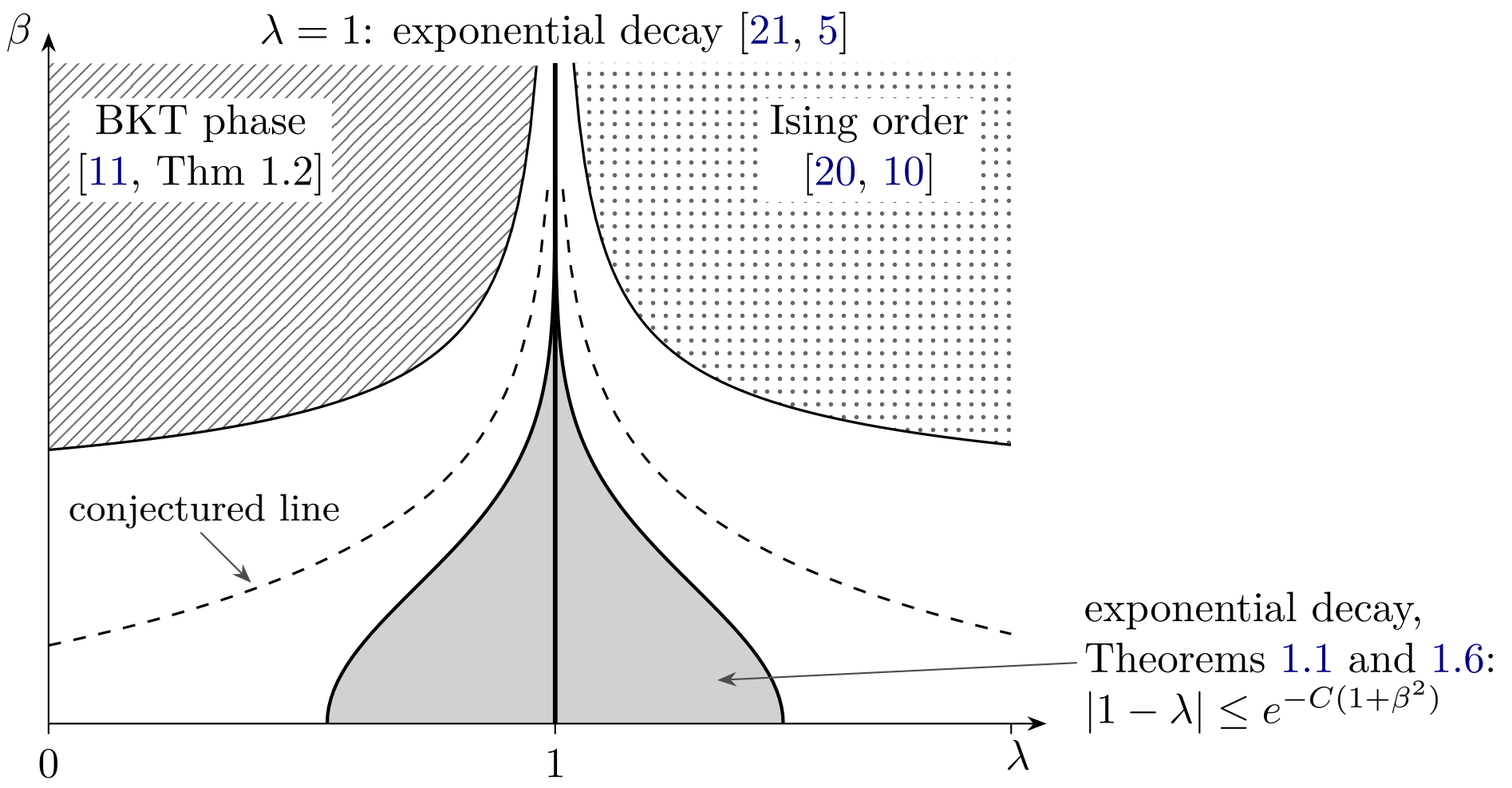
\begin{figure}[tp]
\centering
\begin{tikzpicture}[x=4.2cm,y=0.72cm, >={Stealth[length=1.6mm]}, font=\small]
  \fill[black!18] plot[domain=0:7.6, samples=80] ({1-0.45*exp(-0.2*\x*\x)},\x)
     -- plot[domain=7.6:0, samples=80] ({1+0.45*exp(-0.2*\x*\x)},\x) -- cycle;
  \fill[pattern=north east lines, pattern color=black!55]
     plot[domain=0.0:0.9636, samples=80] (\x,{2.1+1.05/sqrt(1-\x)}) -- (0.0,7.6) -- cycle;
  \draw[line width=0.6pt] plot[domain=0.0:0.9636, samples=80] (\x,{2.1+1.05/sqrt(1-\x)});
  \fill[pattern=dots, pattern color=black!60]
     plot[domain=1.0364:1.9, samples=80] (\x,{2.1+1.05/sqrt(\x-1)}) -- (1.9,7.6) -- cycle;
  \draw[line width=0.6pt] plot[domain=1.0364:1.9, samples=80] (\x,{2.1+1.05/sqrt(\x-1)});
  \draw[dashed, line width=0.6pt] plot[domain=0.0:0.985, samples=60] (\x,{0.9+1.25*ln(1/(1-\x))});
  \draw[dashed, line width=0.6pt] plot[domain=1.015:1.9, samples=60] (\x,{0.9+1.25*ln(1/(\x-1))});
  \draw[line width=0.8pt] plot[domain=0:7.6, samples=80] ({1-0.45*exp(-0.2*\x*\x)},\x);
  \draw[line width=0.8pt] plot[domain=0:7.6, samples=80] ({1+0.45*exp(-0.2*\x*\x)},\x);
  \draw[line width=1.1pt] (1,0) -- (1,7.6);
  \draw[->] (0,0) -- (1.97,0) node[below left] {$\lambda$};
  \draw[->] (0,0) -- (0,7.95) node[left] {$\beta$};
  \foreach \t in {0,1}{\draw (\t,0) -- (\t,-0.12) node[below] {$\t$};}
  \draw (1.9,0) -- (1.9,-0.12);
  \node[fill=white, inner sep=1.5pt, align=center] at (0.30,6.6) {BKT phase\\ \cite[Thm~1.2]{GarbanMontgolfier2026}};
  \node[fill=white, inner sep=1.5pt, align=center] at (1.62,6.6) {Ising order\\ \cite{Malyshev1975,FrohlichLieb1978}};
  \node[anchor=west, align=left, inner sep=1.5pt] (ours) at (2.03,0.7)
     {exponential decay,\\ Theorems~\ref{thm:main} and~\ref{thm:rate}:\\ $|1-\lambda|\le e^{-C(1+\beta^2)}$};
  \draw[->, black!70] (ours.west) -- (1.32,0.5);
  \node[anchor=south, fill=white, inner sep=1.5pt, align=center] at (1.0,7.62) {\small $\lambda=1$: exponential decay~\cite{OpenAI215,DHY2026}};
  \node[anchor=west, fill=white, inner sep=1pt] at (0.03,2.45) {\footnotesize conjectured line};
  \draw[->, black!70] (0.30,2.2) -- (0.40,1.62);
\end{tikzpicture}
\caption{A schematic phase diagram of the $\lambda$-model ($n=3$), not to scale. Along $\lambda=1$ correlations
decay exponentially at every $\beta$~\cite{OpenAI215,DHY2026}. Theorems~\ref{thm:main} and~\ref{thm:rate} give
exponential decay in the grey region around $\lambda=1$, whose half-width at height $\beta$ is at least
$e^{-C(1+\beta^2)}$ (drawn much wider than it is); it contains every point below it at the same $\lambda$,
because the estimates are uniform in strengths up to $\beta$. The hatched region is the BKT phase
of~\cite[Theorem~1.2]{GarbanMontgolfier2026}, whose boundary $\bar\beta(\lambda)$ must therefore stay above the
grey region and tend to infinity as $\lambda\uparrow1$. The dotted region is the Ising phase for $\lambda>1$. The
dashed curves sketch the conjectured transition lines $\beta\approx(4\pi)^{-1}|\log|1-\lambda||$, which the
renormalization-group computation gives on both sides~\cite{HikamiTsuneto1980,IrkhinKatanin1999}; the drawn
shapes of the hatched and dotted boundaries are illustrative only.}
\label{fig:phase}
\end{figure}

\subsection{Idea of the proofs}\label{sec:ideas}

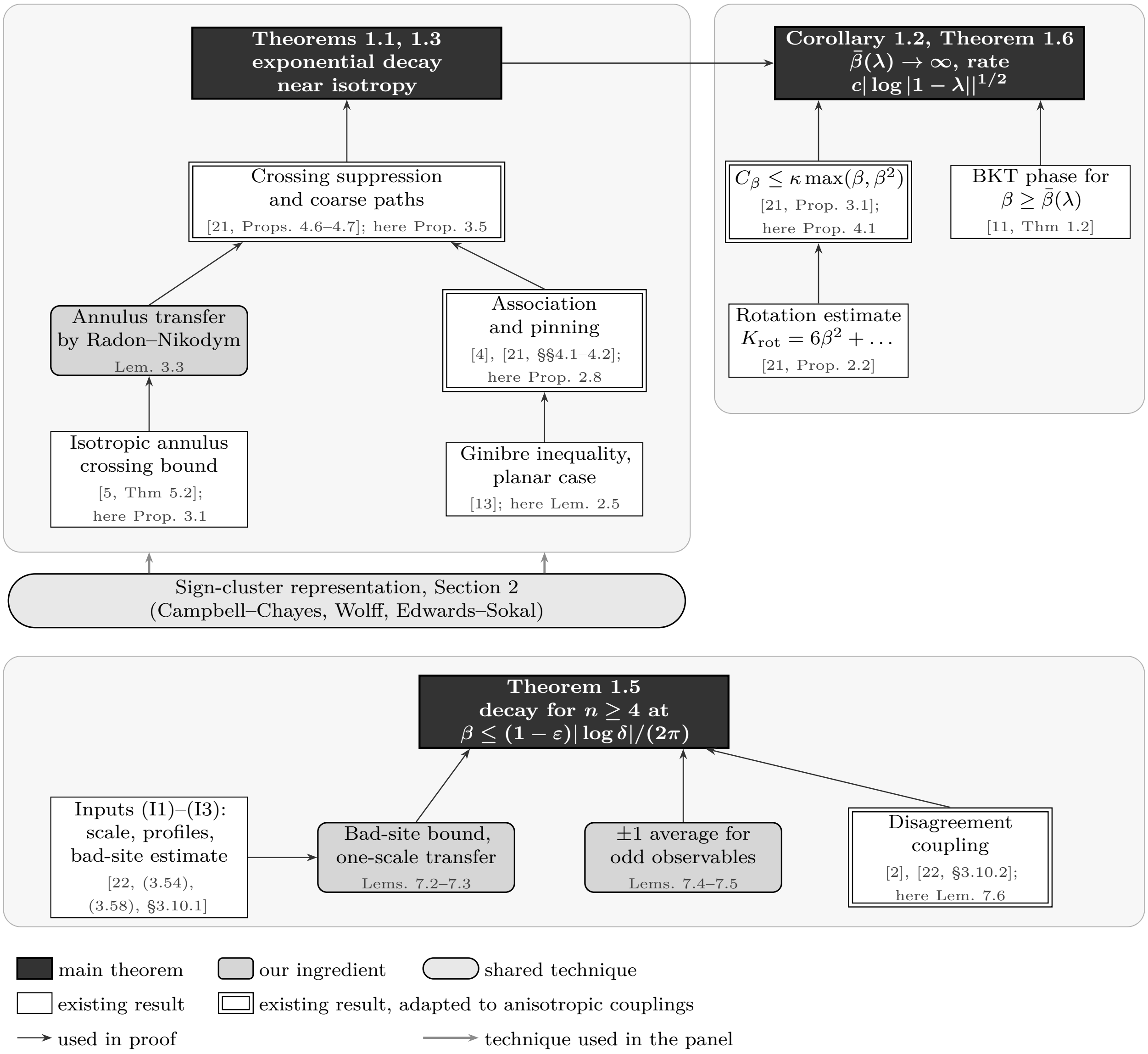
\begin{figure}[tp]
\centering
\newcommand*{\rmloc}[1]{\\[1pt]{\tiny\color{black!75}#1}}%
\begin{NoHyper}%
\begin{tikzpicture}[
  x=1cm, y=1cm, font=\scriptsize,
  execute at begin node={\thickmuskip=5mu\medmuskip=4mu\relax},
  >={Stealth[length=1.5mm,width=1.1mm]},
  thmlook/.style={fill=black!80, draw=black, line width=0.8pt},
  ourslook/.style={rounded corners=1.3mm, fill=black!16, draw=black, line width=0.6pt},
  knownlook/.style={fill=white, draw=black, line width=0.4pt},
  redolook/.style={fill=white, draw=black, line width=0.45pt,
    path picture={\draw[black, line width=0.45pt]
      ([shift={(1.6pt,1.6pt)}]path picture bounding box.south west) rectangle
      ([shift={(-1.6pt,-1.6pt)}]path picture bounding box.north east);}},
  toollook/.style={rounded rectangle, fill=black!9, draw=black, line width=0.6pt},
  box/.style={rectangle, align=center, text width=2.75cm, inner xsep=1.2pt, inner ysep=1.5pt,
    minimum height=0.9cm},
  wide/.style={text width=4.37cm},
  thm/.style={box, wide, thmlook, text=white, font=\scriptsize\bfseries},
  ours/.style={box, ourslook},
  known/.style={box, knownlook},
  redo/.style={box, redolook, inner xsep=2.6pt, inner ysep=2.8pt},
  tool/.style={toollook, align=center, inner sep=2.5pt, minimum height=0.5cm},
  sw/.style={rectangle, minimum width=0.5cm, minimum height=0.3cm, inner sep=0pt},
  lg/.style={anchor=west, inner sep=2pt, text height=1.6ex, text depth=0.3ex},
  use/.style={->, line width=0.5pt, draw=black!80},
  rel/.style={use, dashed},
  band/.style={->, line width=0.9pt, draw=black!45},
  col/.style={rounded corners=2.5mm, fill=black!3, draw=black!30, line width=0.4pt},
]
\draw[col] (0.00,0.85) rectangle (9.90,-7.05);
\draw[col] (10.25,0.85) rectangle (16.45,-5.05);
\node[thm] (T1) at (4.95,0) {Theorems~\ref{thm:main}, \ref{thm:ext}\\ exponential decay\\ near isotropy};
\node[thm] (T2) at (13.35,0) {Corollary~\ref{cor:lambda}, Theorem~\ref{thm:rate}\\ $\bar\beta(\lambda)\to\infty$, rate\\ $c|\log|1-\lambda||^{1/2}$};
\node[redo, wide] (C6) at (4.95,-2.0) {Crossing suppression\\ and coarse paths\rmloc{\cite[Props.~4.6--4.7]{OpenAI215}; here Prop.~\ref{prop:connection}}};
\node[ours]  (RN) at (2.10,-4.0) {Annulus transfer\\ by Radon--Nikodym\rmloc{Lem.~\ref{lem:rn}}};
\node[redo]  (PI) at (7.80,-4.0) {Association\\ and pinning\rmloc{\cite{CampbellChayes1998}, \cite[\S\S4.1--4.2]{OpenAI215}; here Prop.~\ref{prop:pinning}}};
\node[known] (AN) at (2.10,-6.0) {Isotropic annulus\\ crossing bound\rmloc{\cite[Thm~5.2]{DHY2026}; here Prop.~\ref{prop:dhy}}};
\node[known] (GI) at (7.80,-6.0) {Ginibre inequality,\\ planar case\rmloc{\cite{Ginibre1970}; here Lem.~\ref{lem:ginibre}}};
\draw[use] (C6) -- (C6 |- T1.south);
\draw[use] (RN.north) -- (C6.south -| 3.45,0);
\draw[use] (PI.north) -- (C6.south -| 6.45,0);
\draw[use] (AN) -- (RN);
\draw[use] (GI) -- (PI);
\node[redo, text width=2.5cm]  (CB) at (11.75,-2.0) {$C_\beta\le\kappa\max(\beta,\beta^2)$\rmloc{\cite[Prop.~3.1]{OpenAI215}; here Prop.~\ref{prop:cbeta}}};
\node[known, text width=2.5cm] (GD) at (14.95,-2.0) {BKT phase for\\ $\beta\ge\bar\beta(\lambda)$\rmloc{\cite[Thm~1.2]{GarbanMontgolfier2026}}};
\node[known, text width=2.5cm] (K2) at (11.75,-4.0) {Rotation estimate\\ $K_{\rm rot}=6\beta^2+\dots$\rmloc{\cite[Prop.~2.2]{OpenAI215}}};
\draw[use] (CB) -- (CB |- T2.south);
\draw[use] (GD) -- (GD |- T2.south);
\draw[use] (K2) -- (CB);
\draw[use] (T1) -- (T2);
\node[tool, text width=9.3cm] (S1) at (4.95,-7.75) {Sign-cluster representation, Section~\ref{sec:clusters}\\ (Campbell--Chayes, Wolff, Edwards--Sokal)};
\draw[band] (2.10,-7.75 |- S1.north) -- (2.10,-7.05);
\draw[band] (7.80,-7.75 |- S1.north) -- (7.80,-7.05);
\draw[col] (0.00,-8.55) rectangle (16.45,-12.45);
\node[thm] (T3) at (8.225,-9.35) {Theorem~\ref{thm:phys}\\ decay for $n\ge4$ at\\ $\beta\le(1-\varepsilon)|\log\delta|/(2\pi)$};
\node[known] (OI) at (2.10,-11.45) {Inputs (I1)--(I3):\\ scale, profiles,\\ bad-site estimate\rmloc{\cite[(3.54), (3.58), \S3.10.1]{OpenAIO4}}};
\node[ours]  (BS) at (5.95,-11.45) {Bad-site bound,\\ one-scale transfer\rmloc{Lems.~\ref{lem:celltransfer}--\ref{lem:badsite}}};
\node[ours]  (OD) at (9.80,-11.45) {$\pm1$ average for\\ odd observables\rmloc{Lems.~\ref{lem:pm1}--\ref{lem:odd}}};
\node[redo]  (CP) at (13.65,-11.45) {Disagreement coupling\rmloc{\cite{vdBergMaes1994}, \cite[\S3.10.2]{OpenAIO4}; here Lem.~\ref{lem:coupling}}};
\draw[use] (OI) -- (BS);
\draw[use] (BS.north) -- ($(T3.south)+(-1.5,0)$);
\draw[use] (OD.north) -- (OD.north |- T3.south);
\draw[use] (CP.north) -- ($(T3.south)+(1.9,0)$);
\node[sw, thmlook]  (g1) at (0.45,-13.05) {}; \node[lg] at (g1.east) {main theorem};
\node[sw, ourslook, rounded corners=0.6mm] (g2) at (3.35,-13.05) {}; \node[lg] at (g2.east) {our ingredient};
\node[sw, toollook, minimum width=0.8cm] (g3) at (6.45,-13.05) {}; \node[lg] at (g3.east) {shared technique};
\node[sw, knownlook] (g4) at (0.45,-13.55) {}; \node[lg] at (g4.east) {existing result};
\node[sw, redolook]  (g5) at (3.35,-13.55) {}; \node[lg] at (g5.east) {existing result, adapted to anisotropic couplings};
\draw[use]  (0.20,-14.05) -- (0.70,-14.05) node[lg] {used in proof};
\draw[band] (6.05,-14.05) -- (6.85,-14.05) node[lg] {technique used in the panel};
\end{tikzpicture}%
\end{NoHyper}
\caption{How the results are proved. A solid arrow means that its source is used in the proof of its
target; only the main dependencies are drawn. Box styles give provenance; grey labels give sources and
where each item is proved. The adapted steps are listed in Section~\ref{sec:ideas}.}
\label{fig:roadmap}
\end{figure}

The proofs of Theorems~\ref{thm:main}, \ref{thm:ext} and~\ref{thm:rate} share one scheme, which goes back
to Campbell and Chayes~\cite{CampbellChayes1998} and is carried out for the isotropic model
in~\cite[\S4]{OpenAI215}: a sign-cluster representation turns a crossing bound for a single isotropic
square annulus $\{N\le\|v\|_\infty\le2N\}$ into exponential decay. The signs of one spin component form a
ferromagnetic Ising model given the amplitudes, open bonds join equal signs, pinning annular layers to the
pole increases crossing probabilities, and a sum over coarse paths at one fixed scale $N$ gives a positive
mass. This part uses only positive association, and its isotropy enters through a Ginibre-type inequality
for the planar components~\cite[Lemma~4.2]{OpenAI215}. The annulus bound has two suppliers. For
Theorems~\ref{thm:main} and~\ref{thm:ext} it is~\cite[Theorem~5.2]{DHY2026}. Ding, Huang and Yang prove
exponential decay through loop soups and Lupu's isomorphism~\cite{Lupu2016}; their key estimate bounds the
probability that an annulus is crossed by a loop cluster, uniformly over a class of boundary conditions, and
its proof uses spin waves on $S^1$ instead of rotations of $S^2$ about two non-commuting
axes~\cite[Remark~1.5]{DHY2026}. For Theorem~\ref{thm:rate} it is~\cite[Proposition~3.1]{OpenAI215}: if the
outer layer of the annulus is pinned to $e_1$ and its inner layer to $e_1$ rotated by an angle $t$ about an
axis orthogonal to $e_1$, the partition function $Z(t)$ satisfies $\min_tZ(t)/\max_tZ(t)\ge1-C_\beta/\log N$.
This step uses rotations about two non-commuting axes, and it fails for anisotropic couplings; this is why
we use the isotropic input only on one finite annulus.

For anisotropic couplings we keep the sign-cluster part and use the isotropic input on one annulus at one
scale $N$ only. When both layers are pinned to the pole, the bound
of~\cite[Theorem~5.2]{DHY2026} is a bound on the probability that open sign-cluster bonds cross the
isotropic annulus (Proposition~\ref{prop:dhy}). On that finite graph the anisotropic and isotropic
Hamiltonians differ by at most $\beta\delta|E_N|$, so the two partition functions with the same pins agree up
to a factor $e^{\pm\beta\delta|E_N|}$. Choosing $N$ first, as a function of $\beta$, and then
$\delta\approx(\beta N^2)^{-1}$ transfers the crossing bound (Lemma~\ref{lem:rn}). For the second part, the
planar energy $J^j\cos\theta_x\cos\theta_y+J^k\sin\theta_x\sin\theta_y$ is a combination with nonnegative
coefficients of $\cos(\theta_x-\theta_y)$ and $\cos(\theta_x+\theta_y)$. Ginibre's doubling argument handles
both terms in the same way (Lemma~\ref{lem:ginibre}), so amplitude association, pinning and the crossing
identity go through; Section~\ref{sec:clusters} writes these steps out. The scale $N(\beta)$ is where the
price is paid: $\delta(\beta)$ is of order $(\beta N(\beta)^2)^{-1}$. The scale given by~\cite{DHY2026} is not
explicit. With the input of~\cite[Proposition~3.1]{OpenAI215} instead, $\log N(\beta)$ is proportional to
$C_\beta$, and Proposition~\ref{prop:cbeta} shows that $C_\beta$ is at most quadratic in $\beta$; the quadratic
term comes from the coefficient $K_{\mathrm{rot}}(\beta)=6\beta^2+4\beta^{3/2}+\beta$ of the rotation
estimate~\cite[Proposition~2.2]{OpenAI215}. This gives Theorem~\ref{thm:rate}.

Theorem~\ref{thm:phys} is proved differently, following the proof of decay for charged observables of the
isotropic $O(4)$ model in~\cite[\S\S3.9--3.10]{OpenAIO4}: orientation variables on coarse cells of side
$s(\beta)=e^{\pi\beta+O(\sqrt{\beta\log\beta})}$, a test for bad coarse vertices, at which rotating the
orientation can change the cell actions by more than a fixed amount, and a disagreement coupling in the
manner of van den Berg and Maes~\cite{vdBergMaes1994}. The anisotropic and isotropic actions of one coarse
cell differ by at most $8\beta\delta s^2$ (Lemma~\ref{lem:celltransfer}). As long as $\beta\delta s^2$ is small,
two things survive. The isotropic bad-site estimate~\cite[(3.58)]{OpenAIO4}, which holds conditionally on
everything outside a collar, transfers to the anisotropic law through a bounded density ratio on the
collar (Lemma~\ref{lem:badsite}). At good vertices the one-site conditional law of the orientation stays
close to Haar measure. The averaging over common left rotations by all of $S^3$ in~\cite{OpenAIO4} needs
isotropy; we replace it by the symmetry $\sigma\mapsto-\sigma$, which every bilinear interaction keeps and
which suffices for odd observables (Lemmas~\ref{lem:pm1}--\ref{lem:odd}). The condition that $\beta\delta s^2$
be small is the crossover condition with $\xi=s$.

\emph{Adapted results.} Lemmas~\ref{lem:amplitude} and~\ref{lem:joint} and Proposition~\ref{prop:pinning}
adapt~\cite[Lemma~2, Corollaries~1 and~3]{CampbellChayes1998} and~\cite[Lemmas~4.3, 4.4 and
Proposition~4.5]{OpenAI215}, with Lemma~\ref{lem:ginibre} in place of~\cite[Lemma~4.2]{OpenAI215};
Lemma~\ref{lem:crossing} is~\cite[(4.9)]{OpenAI215}; Lemma~\ref{lem:indep} and
Proposition~\ref{prop:connection} adapt~\cite[Propositions~4.6 and~4.7]{OpenAI215};
Proposition~\ref{prop:dhy} translates~\cite[Theorem~5.2]{DHY2026} into a bound on sign-cluster crossings;
Proposition~\ref{prop:cbeta} extracts the $\beta$-dependence of the constant in~\cite[Proposition~3.1]{OpenAI215}
from its proof; Theorem~\ref{thm:ext}(b) uses the conditional projection
of~\cite[Proposition~2.4]{AruGarbanSepulveda2025} in the block form of~\cite[Lemma~4.8]{OpenAI215};
Section~\ref{sec:phys} adapts~\cite[\S\S3.9--3.10]{OpenAIO4}, with the credits given there.

Section~\ref{sec:clusters} sets up sign clusters for anisotropic couplings; Section~\ref{sec:main} proves
Theorem~\ref{thm:main}, Section~\ref{sec:rate} Theorem~\ref{thm:rate}, Section~\ref{sec:ext}
Theorem~\ref{thm:ext}, Section~\ref{sec:cor} the corollaries, and Section~\ref{sec:phys}
Theorem~\ref{thm:phys}.

\section{Sign clusters for anisotropic couplings}\label{sec:clusters}

This section writes out, for anisotropic couplings, the sign-cluster representation of Campbell and
Chayes~\cite{CampbellChayes1998}, which builds on Wolff's embedding~\cite{Wolff1989} and the
Edwards--Sokal coupling~\cite{EdwardsSokal1988}, and the association and pinning arguments
of~\cite[\S\S4.1--4.2]{OpenAI215}. Fix $\Jv\in(0,\infty)^3$ and a coordinate $i$. By permuting coordinates
(which preserves the reference measure) we may assume $i=1$ and $J^2\ge J^3$.

\begin{lemma}[decomposition]\label{lem:decomp}
Write, as in~\cite[(4.1)]{OpenAI215},
\[
\sigma_x=(r_x\tau_x,\ q_x\cos\theta_x,\ q_x\sin\theta_x),\qquad q_x=(1-r_x^2)^{1/2}.
\]
Under the reference measure $r_x,\tau_x,\theta_x$ are independent and uniform on $[0,1]$, $\{\pm1\}$ and
$\R/2\pi\Z$. The weight of an edge is $b_{xy}\bigl(J^1r_xr_y\tau_x\tau_y+q_xq_yT_{xy}\bigr)$ with
\begin{equation}\label{eq:Txy}
T_{xy}=J^2\cos\theta_x\cos\theta_y+J^3\sin\theta_x\sin\theta_y=\alpha\cos(\theta_x-\theta_y)+\gamma\cos(\theta_x+\theta_y),
\end{equation}
$\alpha=\frac12(J^2+J^3)>0$, $\gamma=\frac12(J^2-J^3)\ge0$. Conditionally on $r$, the signs form a
ferromagnetic Ising system with couplings $K_{xy}=b_{xy}J^1r_xr_y$ and the angles an independent planar
system with couplings $L_{xy}=b_{xy}q_xq_y$ and edge energies $T_{xy}$; the amplitude law has density
$\rho(r)\propto Z_I(K(r))Z_P(L(r))$ on $[0,1]^V$, which is positive and continuous.
\end{lemma}
\begin{proof}
By Archimedes' theorem the first coordinate $r\tau$ of a uniform point of $S^2$ is uniform on $[-1,1]$ and
independent of the azimuth $\theta$. The rest is algebra.
\end{proof}

\begin{lemma}[bonds and the connection identity]\label{lem:bonds}
Given the spins, sample independent bonds with $\Pp(\eta_e=1\mid\sigma)=p_e\1\{\tau_x=\tau_y\}$,
$p_e=1-e^{-2K_e}$. Conditionally on $(r,\theta,\eta)$, the signs are independent fair signs, constant on open
clusters; conditionally on $r$, the bonds are independent of $\theta$ and have law proportional to
$2^{k(\eta)}\prod_ep_e^{\eta_e}(1-p_e)^{1-\eta_e}$, where $k(\eta)$ is the number of open clusters, singletons
included. Consequently
\begin{equation}\label{eq:spin-connection}
\langle\sigma_x^1\sigma_y^1\rangle_{G,b,\Jv}=\E\bigl[r_xr_y\1\{x\lr y\}\bigr]\in\bigl[0,\Pp(x\lr y)\bigr].
\end{equation}
\end{lemma}
\begin{proof}
This is the Edwards--Sokal coupling~\cite{EdwardsSokal1988} for the embedded Ising
model~\cite{Wolff1989}; see~\cite[Corollary~4, (7)]{CampbellChayes1998} and~\cite[(4.4)--(4.6)]{OpenAI215}.
With $p=1-e^{-2K}$ one has $e^{K\tau\tau'}=e^K[(1-p)+p\1\{\tau=\tau'\}]$. Expanding the product over edges,
the joint density of $(r,\tau,\theta,\eta)$ is proportional to
$\prod_ee^{K_e}p_e^{\eta_e}(1-p_e)^{1-\eta_e}\prod_{e:\eta_e=1}\1\{\tau_x=\tau_y\}\cdot\exp(\sum_eL_eT_e(\theta))$.
Its dependence on $\tau$ given $(r,\theta,\eta)$ is the indicator that $\tau$ is constant on open clusters,
summing over $\eta$ recovers the Gibbs density, and the conditional law of $\eta$ given $\sigma$ is the stated
kernel. Summing over $\tau$ gives the law of $\eta$ given $r$. Hence $\E[\tau_x\tau_y\mid r,\theta,\eta]=\1\{x\lr y\}$
and $\langle\sigma_x^1\sigma_y^1\rangle=\E[r_xr_y\tau_x\tau_y]$.
\end{proof}

\begin{lemma}[FKG inequality]\label{lem:fkg}
Let $X=\prod_{k=1}^mX_k$, where each $X_k$ is a compact interval with Lebesgue measure or a finite chain
with counting measure, and let $w>0$ be continuous and log-supermodular:
$w(u\vee v)w(u\wedge v)\ge w(u)w(v)$. Then the law $\mu\propto w$ is associated: $\Cov(F,H)\ge0$ for bounded
measurable coordinatewise increasing $F,H$. Moreover the conditional law of $(x_1,\dots,x_{m-1})$ given
$x_m$ is stochastically increasing in $x_m$.
\end{lemma}
\begin{proof}
This is the FKG inequality~\cite{FortuinKasteleynGinibre1971,Preston1974}; we give the standard induction,
as in~\cite[Lemma~4.1]{OpenAI215}. For $m=1$, with $X'$ an independent copy,
$2\Cov(F,H)=\E[(F(X)-F(X'))(H(X)-H(X'))]\ge0$, since both differences have the sign of $X-X'$. For $m>1$
write $x=(u,t)$. The conditional density $w(\cdot,t)$ is positive, continuous and log-supermodular in $u$, so
the conditional law $\mu_t$ is associated by induction. For $t'\ge t$ and $u\le v$, log-supermodularity
applied to $(u,t')$ and $(v,t)$ gives $w(v,t')w(u,t)\ge w(u,t')w(v,t)$, so $g(u):=w(u,t')/w(u,t)$ is
increasing and bounded. Then $\E_{t'}[F]=\E_t[Fg]/\E_t[g]\ge\E_t[F]$ for increasing $F$, by association of
$\mu_t$; this is the second assertion, and it makes $t\mapsto\E[F\mid t]$ increasing. Finally
$\Cov(F,H)=\E[\Cov(F,H\mid t)]+\Cov(\E[F\mid t],\E[H\mid t])\ge0$, the first term by induction and the second by
the case $m=1$.
\end{proof}

\begin{lemma}[Griffiths' inequalities]\label{lem:griffiths}
For an Ising model on a finite graph with free boundary conditions and couplings $K_e\ge0$, every
$\langle\tau_x\tau_y\rangle\ge0$ and every $\Cov(\tau_x\tau_y,\tau_u\tau_v)\ge0$.
\end{lemma}
\begin{proof}
These are Griffiths' inequalities in the form of Kelly and Sherman~\cite{KellySherman1968} and
Ginibre~\cite{Ginibre1970}; we give Ginibre's doubling proof, as in~\cite[Lemma~4.2]{OpenAI215}. The means
are nonnegative because, after expanding the exponential, the reference moments of sign monomials are $0$
or $1$. For the covariances take two independent copies $d,d'$ and put $u=(d+d')/2$, $v=(d-d')/2$ at each
site. The pair $(u,v)$ takes the values $(\pm1,0)$ and $(0,\pm1)$ with probability $\frac14$ each, so every
monomial in $(u,v)$ at one site has nonnegative expectation. Since $d_xd_y+d'_xd'_y=2(u_xu_y+v_xv_y)$ and
$d_xd_y-d'_xd'_y=2(u_xv_y+v_xu_y)$, the quantity $2\Cov(\tau_x\tau_y,\tau_u\tau_v)$ is a ratio whose numerator,
after expanding $\exp(2\sum_eK_e(u_xu_y+v_xv_y))$, is a sum of nonnegative coefficients times products over
sites of nonnegative moments.
\end{proof}

\begin{lemma}[planar Ginibre inequality]\label{lem:ginibre}
Consider a planar system on a finite graph with density proportional to
$\exp(\sum_aL_af_a(\theta))$, where $L_a\ge0$ and each $f_a$ is $\cos(n_a\cdot\theta)$ for an integer
vector $n_a$. Then every $\langle f_a\rangle\ge0$ and every $\Cov(f_a,f_b)\ge0$. In particular this holds for
the edge energies~\eqref{eq:Txy} with $L_e\ge0$, and also with additional site terms $M_x\cos(2\theta_x)$,
$M_x\ge0$.
\end{lemma}
This is a case of Ginibre's inequalities~\cite{Ginibre1970}; in this form it also follows from Dunlop's
inequality as restated in~\cite[Theorem~A.2 and (A.1)--(A.2)]{GarbanMontgolfier2026}. We include the
short proof, which follows~\cite[proof of Lemma~4.2]{OpenAI215}.
\begin{proof}
Expanding the exponential, products of cosines of integer combinations are nonnegative combinations of
such cosines, whose Haar integrals are nonnegative; this gives $\langle f_a\rangle\ge0$. For covariances,
take two independent copies $\theta,\theta'$ and substitute $\theta=\varphi+\psi$, $\theta'=\varphi-\psi$; the map
$(\varphi,\psi)\mapsto(\varphi+\psi,\varphi-\psi)$ is a surjective homomorphism of tori, so it preserves Haar
probability~\cite[proof of Lemma~4.2]{OpenAI215}. For every integer vector $n$,
\[
\cos(n\cdot\theta)+\cos(n\cdot\theta')=2\cos(n\cdot\varphi)\cos(n\cdot\psi),\qquad
\cos(n\cdot\theta)-\cos(n\cdot\theta')=-2\sin(n\cdot\varphi)\sin(n\cdot\psi).
\]
So the doubled density is $\exp(2\sum_aL_a\cos(n_a\cdot\varphi)\cos(n_a\cdot\psi))$ and
$(f_a-f'_a)(f_b-f'_b)=4\sin(n_a\cdot\varphi)\sin(n_b\cdot\varphi)\sin(n_a\cdot\psi)\sin(n_b\cdot\psi)$. Expanding the
exponential, $2\Cov(f_a,f_b)=\E[(f_a-f'_a)(f_b-f'_b)]$ is a sum of nonnegative coefficients times
$\bigl(\int F\,d\varphi\bigr)^2$ for real trigonometric polynomials $F$. Absolute convergence on the finite
torus justifies the expansions. For~\eqref{eq:Txy} write $L_eT_e=L_e\alpha\cos(n_e^-\cdot\theta)+L_e\gamma
\cos(n_e^+\cdot\theta)$ with $n_e^\pm=e_x\pm e_y$, and use bilinearity of the covariance.
\end{proof}

\begin{lemma}[amplitude association]\label{lem:amplitude}
The amplitude law with density $\rho$ is associated.
\end{lemma}
\begin{proof}
As in~\cite[Lemma~2]{CampbellChayes1998} and~\cite[Lemma~4.3]{OpenAI215}. For a partition function
$Z(c)=\int\exp(\sum_ac_af_a)$ one has $\partial^2\log Z/\partial c_a\partial c_b=\Cov(f_a,f_b)$, so for distinct
sites $x,y$,
\[
\partial_x\partial_y\log Z(c(r))=\sum_a\langle f_a\rangle\,\partial_x\partial_yc_a+\sum_{a,b}\Cov(f_a,f_b)\,\partial_xc_a\,\partial_yc_b.
\]
For $Z_I$, the coefficients $K_{xy}=b_{xy}J^1r_xr_y$ are nondecreasing with $\partial_x\partial_yK_{xy}=b_{xy}J^1\ge0$,
and Lemma~\ref{lem:griffiths} gives nonnegative means and covariances. For $Z_P$, the coefficients
$L_{xy}=b_{xy}q_xq_y$ are nonincreasing, so products of two first derivatives are nonnegative,
$\partial_x\partial_yL_{xy}=b_{xy}r_xr_y/(q_xq_y)\ge0$, and Lemma~\ref{lem:ginibre} gives nonnegative means and
covariances. So $\log\rho$ is $C^2$ on $[0,1)^V$ with nonnegative mixed derivatives, hence supermodular
there; by continuity $\rho$ is log-supermodular on $[0,1]^V$, and Lemma~\ref{lem:fkg} applies.
\end{proof}

\begin{lemma}[joint association]\label{lem:joint}
The joint law of $(r,\eta)$ is associated.
\end{lemma}
\begin{proof}
As in~\cite[Corollary~1]{CampbellChayes1998} and~\cite[Lemma~4.4]{OpenAI215}. Given $r$, the bond law of
Lemma~\ref{lem:bonds} has weight $2^{k(\eta)}\prod_ep_e^{\eta_e}(1-p_e)^{1-\eta_e}$, and $k$ is supermodular
($|V|$ minus the rank of a graphic matroid), so this weight is log-supermodular on $\{0,1\}^E$ (edges with
$p_e=0$ are closed almost surely; restrict to the others). By Lemma~\ref{lem:fkg} the conditional bond law
is associated. Raising $p_e$ tilts it by a factor increasing in $\eta_e$, so by association the bond law is
stochastically increasing in each $p_e$, and $p_e=1-\exp(-2b_eJ^1r_xr_y)$ is nondecreasing in $r$ since
$J^1>0$. For bounded increasing $F(r,\eta)$ the map $r\mapsto\E[F(r,\eta)\mid r]$ is therefore increasing, and
$\Cov(F,H)=\E[\Cov(F,H\mid r)]+\Cov(\E[F\mid r],\E[H\mid r])\ge0$ by conditional association and
Lemma~\ref{lem:amplitude}.
\end{proof}

\begin{proposition}[pinning]\label{prop:pinning}
Pinning any set $B$ of spins to $s_*=e_1$, with the same interactions and with bonds sampled by the rule of
Lemma~\ref{lem:bonds}, cannot decrease the probability of an increasing bond event.
\end{proposition}
\begin{proof}
This is~\cite[Proposition~4.5]{OpenAI215}. For $0<\varepsilon<1$ let
$C_\varepsilon=\{\sigma_x^1\ge1-\varepsilon\text{ for all }x\in B\}=\{r_x\ge1-\varepsilon,\ \tau_x=1\text{ for all }x\in B\}$. By
Lemma~\ref{lem:bonds}, $\Pp(C_\varepsilon\mid r,\theta,\eta)=\1\{r_x\ge1-\varepsilon\ \forall x\in B\}\,2^{-k_B(\eta)}=:T_\varepsilon(r,\eta)$,
where $k_B$ is the number of open clusters meeting $B$. Opening a bond cannot increase $k_B$, so
$T_\varepsilon$ is increasing, and for an increasing bond event $A$ Lemma~\ref{lem:joint} gives
$\Pp(A\mid C_\varepsilon)=\E[\1_AT_\varepsilon]/\E[T_\varepsilon]\ge\Pp(A)$. As $\varepsilon\downarrow0$ the law of $\sigma$ given
$C_\varepsilon$ converges weakly to the pinned law, and the bond kernel
$\Pp(\eta_e=1\mid\sigma)=1-\exp(-2b_eJ^1\max\{\sigma_x^1\sigma_y^1,0\})$ is continuous in $\sigma$; so
$\Pp(A\mid C_\varepsilon)$ converges to the pinned probability of $A$.
\end{proof}

For an integer $N\ge2$ and $z\in\Z^2$ let $D_N(z)=\{v:N\le\|v-z\|_\infty\le2N\}$, with inner layer
$\{\|v-z\|_\infty=N\}$ and outer layer $\{\|v-z\|_\infty=2N\}$. An \emph{annulus graph} is a subgraph $H$ of the
nearest-neighbour graph on $D_N(z)$ with strengths $b\in[0,\beta]^{E(H)}$; its \emph{pinned} measure fixes every
vertex of $H$ on either layer at $s_*$ and keeps all edge interactions, edges between pinned vertices
included, and its bonds are sampled by the rule of Lemma~\ref{lem:bonds}. The annulus is \emph{crossed} if an
open path in $H$ joins the two layers.

\begin{lemma}[crossing identity]\label{lem:crossing}
For an annulus graph and any $\Jv\in(0,\infty)^3$, let $Z_\Jv(0)$ be its partition function with both layers
pinned to $s_*$, and $Z_\Jv(\pi)$ the same with the inner layer pinned to $-s_*$. Then
\begin{equation}\label{eq:crossing-identity}
\Pp^{\mathrm{pinned}}_\Jv(\text{the annulus is crossed})=1-\frac{Z_\Jv(\pi)}{Z_\Jv(0)}.
\end{equation}
\end{lemma}
\begin{proof}
This is~\cite[(4.9)]{OpenAI215}. Pinned sites have $r=1$ and $q=0$ for either sign, so their planar factors
and all couplings $K,L$ are the same in $Z_\Jv(0)$ and $Z_\Jv(\pi)$. Expand every Ising factor as in the proof
of Lemma~\ref{lem:bonds} and sum over the unpinned signs: for each $\eta$ the sum equals
$2^{k_0(\eta)-|I|}$ times the indicator that every open cluster carries a single prescribed sign, where $I$ is
the set of unpinned vertices and $k_0$ the number of open clusters avoiding the layers. With all pins $+$ the
indicator is $1$; with the inner layer $-$ it is the indicator that no open cluster meets both layers. All
other factors coincide at every $(r_I,\theta_I)$. If a layer has no vertex, both sides vanish.
\end{proof}

\section{The annulus input and the proof of Theorem~\ref{thm:main}}\label{sec:main}

Let $\varepsilon_N(\beta)$ be the supremum, over $z\in\Z^2$, annulus graphs $H$ on $D_N(z)$ and strengths in
$[0,\beta]^{E(H)}$, of the pinned crossing probability of the isotropic model $\Jv=(1,1,1)$. By
Lemma~\ref{lem:crossing}, $1-\varepsilon_N(\beta)$ is the infimum of $Z_{(1,1,1)}(\pi)/Z_{(1,1,1)}(0)$.

\begin{proposition}[the isotropic annulus input]\label{prop:dhy}
Let $\beta>0$, $\beta_0=\max\{1,\beta/2\}$, and let $M_0=M_0(\beta_0,3,4^{-26})$ be the constant
of~\cite[Theorem~5.2]{DHY2026} for three spin components and their parameter $c_2=4^{-26}$. Then
$\varepsilon_N(\beta)\le4^{-26}$ for every integer $N\ge\max\{2,M_0\}$.
\end{proposition}
\begin{proof}
Translate so that $z=0$, and fix an annulus graph $H$ with strengths $b$.

\emph{Step 1: the full annulus with positive strengths suffices.} Let $F$ be the full nearest-neighbour graph on
$D_N$, and put $\tilde b_e=b_e$ on $E(H)$ and $\tilde b_e=0$ on $E(F)\setminus E(H)$. In the pinned model on
$(F,\tilde b)$, bonds on edges with $\tilde b_e=0$ are closed almost surely, unpinned vertices outside $V(H)$ are
independent uniform spins, and pinned ones outside $V(H)$ only contribute constants; so the law of
$(\sigma|_{V(H)},\eta|_{E(H)})$ is that of the pinned model on $(H,b)$, and the crossing events agree. The
crossing probability is a ratio of integrals over the compact space of unpinned spins of jointly continuous
bounded functions of $(\tilde b,\sigma)$ (the bond kernel is $1-\exp(-2\tilde b_e\max\{\sigma_x^1\sigma_y^1,0\})$), so it
is continuous in $\tilde b\in[0,\beta]^{E(F)}$. It therefore suffices to bound it for
$\tilde b\in(0,\beta]^{E(F)}$.

\emph{Step 2: dictionary.} For $\tilde b\in(0,\beta]^{E(F)}$ put $\hat J_e:=\tilde b_e/\beta\in(0,1]$ (the couplings of~\cite{DHY2026}) and
$\beta_v:=\beta/2\le\beta_0$ for every $v\in D_N$. The spin measure~\cite[(2.3)]{DHY2026} on $F$ with couplings
$\hat J_e$, local time condition $(\beta_v)$ and spin boundary condition $e_1$ on both layers has weight
$\exp(\sum_e2\hat J_e\sqrt{\beta_u\beta_v}\,\sigma_u\cdot\sigma_v)=\exp(\sum_e\tilde b_e\sigma_u\cdot\sigma_v)$: it is our pinned
isotropic measure.

\emph{Step 3: the loop representation.} The spin boundary condition has first coordinate $+1$ and the other
coordinates $0$ on both layers. By~\cite[(2.15)]{DHY2026} and the proof of~\cite[Theorem~1.1]{DHY2026}, where
the three-component version is used, the loop measure coupled to it is $\nu^\xi_{F,\beta}$ with the local time
boundary condition $\xi$ that has $x$-local time $\beta/2$ and $y$- and $z$-local times $0$ on the layers,
wired in every copy; since the $x$- and $y$-local times sum to $\beta_v$ and the $z$-local time vanishes, this
is an $xy$ boundary condition in the sense of~\cite[\S2.2.2]{DHY2026}. Under the coupling the field is
$\varphi^x_v=\sqrt{2\beta_v}\,\sigma_v^1$~\cite[(1.6), (2.1)]{DHY2026}. After conditioning on the $z$-copy the $x$-
and $y$-copies form the loop representation of an $XY$ model~\cite[Remark~3.7]{DHY2026},
and~\cite[Lemma~3.2]{DHY2026} gives: conditionally on the signed fields, the cable indicators are independent
with
\[
\Pp(\hat\omega^x_{uv}=1\mid\varphi)=\1\{\varphi^x_u\varphi^x_v>0\}\bigl(1-e^{-2\hat J_{uv}|\varphi^x_u\varphi^x_v|}\bigr)
=\1\{\tau_u=\tau_v\}\bigl(1-e^{-2\tilde b_{uv}r_ur_v}\bigr),
\]
the bond kernel of Lemma~\ref{lem:bonds} at $J^1=1$; and connections by positive $x$-local time through
complete cables are open-bond connections. So the pinned crossing probability is the $\nu^\xi_{F,\beta}$
probability that the two layers are joined by a path of positive $x$-local time. Such a path contains a
sub-path from its last visit to the inner layer to the next visit to the outer layer, and since $N\ge2$
this sub-path uses no edge between two layer vertices; so neither the cables between boundary vertices nor
the boundary wiring, which~\cite{DHY2026} exclude from connections, matter.

\emph{Step 4.} With three components, local times at most $\beta_0$, couplings $\hat J\in(0,1]^{E(F)}$ and the $xy$
boundary condition $\xi$ on the boundary layers of $\Lambda_{N,2N}=D_N$, \cite[Theorem~5.2]{DHY2026} with
$M=N\ge M_0$ bounds the probability of an $xy$-crossing by $4^{-26}$, and an $x$-crossing is an
$xy$-crossing.
\end{proof}

\begin{remark}\label{rem:dhy}
(1) The bound of~\cite[Theorem~5.2]{DHY2026} is stated with $M_0$ depending only on $\beta_0$, the number of
components and $c_2$, for couplings normalized by $\max_e\hat J_e\le1$~\cite[\S2.1]{DHY2026}; the applications
in~\cite{DHY2026} take $\hat J\equiv1$. We checked that in~\cite[\S\S2--5]{DHY2026} the couplings enter
through exact representations and through the upper bound $2\hat J_e\sqrt{\beta_u\beta_v}\le2\beta_0$, and that no
lower bound on $\hat J_e$ is used; a few displays are written for $\hat J\equiv1$, and with $\hat J_e$ restored they hold for all $\hat J_e\le1$. The
cable of an edge $e$ is to be read with length $1/\hat J_e$, the reading under which~\cite[Lemma~3.2]{DHY2026} holds
as stated. Step~1 removes the need for zero couplings. We use the statements of~\cite{DHY2026} without
re-proving them.
(2) We read the annulus $\Lambda_{M,2M}$ of~\cite[Theorem~5.2]{DHY2026}, which carries the boundary condition
on its two boundary layers, as $D_M$, as in its application~\cite[(5.4)]{DHY2026}. Under any reading with the
pinned layers at sup-radii $a\in[N-1,N+1]$ and $b'\in[2N,2N+1]$, the arguments below run with $D_N$ replaced by
$\{a\le\|v-z\|_\infty\le b'\}$ and $N\ge3$, at the price of replacing $82$ by $100$ in~\eqref{eq:delta} and $124$
by $150$ in Section~\ref{sec:ext}.
\end{remark}

\begin{lemma}[annulus transfer]\label{lem:rn}
If $\Jv\in[1-\delta,1+\delta]^3$ with $\delta\le\frac12$, then for every annulus graph on $D_N(z)$, $N\ge2$, with
strengths in $[0,\beta]$,
\[
\Pp^{\mathrm{pinned}}_\Jv(\text{crossing})\le\varepsilon_N(\beta)+82\beta\delta N^2.
\]
\end{lemma}
\begin{proof}
For both pin choices the Hamiltonians differ pointwise by
\[
\Bigl|\sum_eb_e\sum_i(J^i-1)\sigma_x^i\sigma_y^i\Bigr|\le\beta\delta\sum_e\sum_i|\sigma_x^i||\sigma_y^i|\le\beta\delta|E_N|,
\]
by Cauchy--Schwarz, including edges between pinned sites, where $E_N$ is the edge set of the annulus graph.
Hence $Z_\Jv/Z_{(1,1,1)}\in[e^{-\beta\delta|E_N|},e^{\beta\delta|E_N|}]$ and
$Z_\Jv(\pi)/Z_\Jv(0)\ge e^{-2\beta\delta|E_N|}(1-\varepsilon_N(\beta))$. With~\eqref{eq:crossing-identity},
$1-e^{-x}(1-a)\le x+a$ and $|E_N|\le2|D_N|\le2(4N+1)^2\le41N^2$, the claim follows.
\end{proof}

For $N\ge2$ put
\begin{equation}\label{eq:delta}
\delta_N(\beta)=\min\Bigl\{\frac12,\ \frac{4^{-26}}{82\beta N^2}\Bigr\},\qquad p:=2\cdot4^{-26}=2^{-51}.
\end{equation}
If $\varepsilon_N(\beta)\le4^{-26}$ and $\Jv\in[1-\delta_N(\beta),1+\delta_N(\beta)]^3$, Lemma~\ref{lem:rn} bounds every pinned
crossing probability by $p$.

\begin{lemma}[independent crossings]\label{lem:indep}
Let $N\ge2$ with $\varepsilon_N(\beta)\le4^{-26}$, let $\Jv\in[1-\delta_N(\beta),1+\delta_N(\beta)]^3$ with $J^2\ge J^3$, $G$
finite and $b\in[0,\beta]^E$, and let $z_1,\dots,z_k\in\Z^2$ have pairwise disjoint boxes
$\{\|v-z_j\|_\infty\le2N\}$. Under the free law, the probability that every annulus $D_N(z_j)$ is crossed by an
open path of $G$ inside $D_N(z_j)$ is at most $p^k$.
\end{lemma}
\begin{proof}
This follows~\cite[Proposition~4.6]{OpenAI215}. The intersection of the crossing events is increasing.
Pin to $s_*$ every vertex of $G$ on a layer of one of the annuli; by Proposition~\ref{prop:pinning} this does
not decrease the probability. Let $I_j$ be the set of vertices of $G$ with $N<\|v-z_j\|_\infty<2N$. Along an
edge $\|v-z_j\|_\infty$ changes by at most $1$, so every edge of $G$ with an endpoint in $I_j$ has its other
endpoint in $I_j$ or on a layer of annulus $j$; the $I_j$ are disjoint, and every edge meeting the remaining
unpinned vertices has no endpoint in $\bigcup_jI_j$. With the pins fixed, the Gibbs density therefore
factorizes: the $\sigma|_{I_j}$ are independent, and $\sigma|_{I_j}$ has the pinned law of the annulus graph
$H_j=G\cap D_N(z_j)$. The crossing of annulus $j$ depends only on the bonds of $E(H_j)$, whose conditional law
given $\sigma$ depends only on $\sigma|_{D_N(z_j)}$, and the sets $E(H_j)$ are disjoint. So the crossing events
are independent, each of probability at most $p$.
\end{proof}

\begin{proposition}[connections]\label{prop:connection}
Let $N\ge2$ with $\varepsilon_N(\beta)\le4^{-26}$, and put $A_0=1/(4(1-2^{-1/25}))$ and $m_0=(\log2)/(25\sqrt2N)$. For
all $G$, $b\in[0,\beta]^E$ and $\Jv\in[1-\delta_N(\beta),1+\delta_N(\beta)]^3$ with $J^2\ge J^3$, the bond law
satisfies $\Pp(x\lr y)\le A_0e^{-m_0\|x-y\|_2}$.
\end{proposition}
\begin{proof}
This follows~\cite[Proposition~4.7]{OpenAI215}; the lattice-animal sum in the proof
of~\cite[Theorem~5.1]{DHY2026} has the same shape. Translate so that $x=0$ and let $d=\|y\|_\infty$. If $d\le4N$,
then $\|y\|_2\le4\sqrt2N$ and $A_0e^{-m_0\|y\|_2}\ge A_02^{-4/25}\ge1$. Let $d>4N$. Partition $\Z^2$ into the boxes
$Q_w=w+\{0,\dots,N-1\}^2$, $w\in N\Z^2$. An open path from $0$ to $y$ visits a sequence of boxes, consecutive
ones adjacent, and loop erasure gives a simple coarse path $w^0,\dots,w^k$ of visited boxes with
$Q_{w^0}\ni0$, $Q_{w^k}\ni y$ and $k\ge d/N-1$. For each $j$ take a visited vertex $v\in Q_{w^j}$, so
$\|v-w^j\|_\infty<N$; one of $0,y$ is at sup-distance more than $2N$ from $w^j$, since otherwise $d\le4N$. Along the
open path from $v$ to that endpoint, $\|\cdot-w^j\|_\infty$ moves by unit steps from below $N$ to above $2N$, and
the part between the last visit of the inner layer before the first visit of the outer layer stays in
$D_N(w^j)$: the annulus $D_N(w^j)$ is crossed. One of the $25$ residue classes of $w^j/N$ modulo $5$ contains at
least $(k+1)/25$ of the distinct $w^j$; two distinct centres in one class differ by at least $5N$ in some
coordinate, so their closed $2N$-boxes are disjoint. Choosing such a class deterministically for each coarse
path, Lemma~\ref{lem:indep} bounds the probability that a given coarse path occurs by $p^{(k+1)/25}$. There are
at most $4^k$ simple coarse paths of length $k$ from $Q_{w^0}$. With $\alpha=4p^{1/25}=2^{-1/25}$,
\[
\Pp(0\lr y)\le\sum_{k\ge d/N-1}4^kp^{(k+1)/25}=p^{1/25}\sum_{k\ge d/N-1}\alpha^k\le\frac{p^{1/25}\alpha^{d/N-1}}{1-\alpha}
=A_02^{-d/(25N)}\le A_0e^{-m_0\|y\|_2},
\]
using $p^{1/25}/\alpha=\frac14$ and $d\ge\|y\|_2/\sqrt2$.
\end{proof}

\begin{proof}[Proof of Theorem~\ref{thm:main}]
Let $N(\beta)=\max\{2,\lceil M_0\rceil\}$ with $M_0$ as in Proposition~\ref{prop:dhy}, so that
$\varepsilon_{N(\beta)}(\beta)\le4^{-26}$, and put $\delta(\beta)=\delta_{N(\beta)}(\beta)$, $A=A_0$, $m=m_0$. For $i=1$, relabel
coordinates $2,3$ so that $J^2\ge J^3$; this preserves the reference measure, the box
$[1-\delta,1+\delta]^3$, $\sigma^1$ and the bond rule. By~\eqref{eq:spin-connection} and
Proposition~\ref{prop:connection}, $0\le\langle\sigma_x^1\sigma_y^1\rangle\le\Pp(x\lr y)\le A_0e^{-m_0\|x-y\|_2}$. For
$i=2,3$, permute coordinates; the box is permutation invariant. The cap $\delta\le\frac12$ keeps $J^i>0$, as
Lemmas~\ref{lem:amplitude}--\ref{lem:joint} need.
\end{proof}

The only dependence on $\beta$ is through $N(\beta)$, which~\cite{DHY2026} do not make explicit. A shift of
the scale by one, as in Remark~\ref{rem:dhy}(2), changes only constants.

\section{The dependence on \texorpdfstring{$\beta$}{beta}: proof of Theorem~\ref{thm:rate}}\label{sec:rate}

Let $C_\beta$ be the constant of~\cite[Proposition~3.1]{OpenAI215}: for the isotropic model,
$Z_{(1,1,1)}(\pi)/Z_{(1,1,1)}(0)\ge1-C_\beta/\log N$ for every $N\ge2$, every annulus graph and all strengths in
$[0,\beta]$. By Lemma~\ref{lem:crossing}, $\varepsilon_N(\beta)\le C_\beta/\log N$.

\begin{proposition}\label{prop:cbeta}
There is $\kappa<\infty$, depending only on the fixed test profiles of~\cite[\S3]{OpenAI215}, such that
$C_\beta\le\kappa\max\{\beta,\beta^2\}$.
\end{proposition}
\begin{proof}
We follow the proof of~\cite[Proposition~3.1]{OpenAI215} and record where $\beta$ enters. The Gibbs
input is the gradient estimate $\langle|X((df)g)|^2\rangle\le\beta\|df\|_2^2$ and $R(h,\bar h)\le\beta\|h\|_2^2$
\cite[Lemma~2.1]{OpenAI215}, which is linear in $\beta$; the remaining factors of $\beta$ come from the
pointwise bound $b_e\le\beta$ on local coefficients. The rotation estimate~\cite[Proposition~2.2]{OpenAI215}
states
\[
R(Q,\bar Q)\le K_{\mathrm{rot}}(\beta)\|T\|_{HS}^2+\beta\sum_{i,j}\min\{\|f_i\|_\infty^2\|df_j\|_2^2,\|f_j\|_\infty^2\|df_i\|_2^2\},\qquad
K_{\mathrm{rot}}(\beta)=6\beta^2+4\beta^{3/2}+\beta .
\]
For the test family of~\cite[Lemma~3.2]{OpenAI215}, whose constants depend only on the fixed profiles,
$Q=-i\lambda_N\tilde h$ with $\lambda_N\ge\frac12\sum_{r\le N}r^{-1}\ge\frac12\log N$, and the bracket in
\cite[(3.10)]{OpenAI215} is at most $C(1+\log N)$. Hence, uniformly in the twist,
$R_t(\tilde h,\tilde h)\le4CK_{\mathrm{rot}}(\beta)(1+\log N)/(\log N)^2$, which is the bound
\cite[(3.16)]{OpenAI215} with an explicit $\beta$-dependence. The transfer from $\tilde h$ to $h$
\cite[(3.17)--(3.19)]{OpenAI215} adds terms bounded by $C(\beta^{3/2}/N+\beta^2/N^2+\beta/N^2)$, and the
integration step that ends the proof is free of $\beta$. Since $\log N/N\le1$ and
$\beta^{3/2}\le\max\{\beta,\beta^2\}$, this bounds the quantity estimated in~\cite[(3.20)]{OpenAI215} by
$\kappa'\max\{\beta,\beta^2\}/\log N$, and $C_\beta=\frac{\pi^2}{2}\kappa'\max\{\beta,\beta^2\}$ works.
\end{proof}

\begin{proof}[Proof of Theorem~\ref{thm:rate}]
Take $N=\max\{2,\lceil\exp(4^{26}C_\beta)\rceil\}$, so that $\varepsilon_N(\beta)\le C_\beta/\log N\le4^{-26}$. The proof of
Theorem~\ref{thm:main} then runs with this $N$ in place of $N(\beta)$ and gives Theorem~\ref{thm:main} with
$\delta(\beta)=\delta_N(\beta)$. By Proposition~\ref{prop:cbeta} and $\log N\le4^{26}C_\beta+\log2$,
\[
\log\frac1{\delta(\beta)}\le\max\Bigl\{\log2,\ 2\cdot4^{26}\kappa\max\{\beta,\beta^2\}+\log(82\cdot4^{26}\beta)+2\log2\Bigr\}
\le C(1+\beta^2).
\]
Given $\lambda$ with $|1-\lambda|\le e^{-2C}$, put $L=\log(1/|1-\lambda|)\ge2C$ and
$\beta=(L/C-1)^{1/2}$, so that $C(1+\beta^2)=L$ and $\beta\ge(L/(2C))^{1/2}$. Then
$|1-\lambda|\le\delta(\beta)$, and Corollary~\ref{cor:lambda} applies at every
$\beta'\le(2C)^{-1/2}|\log|1-\lambda||^{1/2}$.
\end{proof}

\begin{remark}
The exponent $\frac12$ is lost in the $\beta^2$ part of $K_{\mathrm{rot}}$, which comes from four terms in which a
local coefficient of size $\beta$ meets a current variance of size $\beta$ and two products of local
coefficients; the physical prediction corresponds to $C_\beta$ of order $\beta$.
\end{remark}

\section{Extensions: proof of Theorem~\ref{thm:ext}}\label{sec:ext}

\begin{proof}[Proof of Theorem~\ref{thm:ext}(a)]
Choose the sign component $i$ and order the other two coordinates $(j,k)$ by $\pi$, so that $J^j\ge J^k$ and
$c_{x,j}\ge c_{x,k}$. The single-ion terms become
$c_{x,i}r_x^2+\frac12q_x^2(c_{x,j}+c_{x,k})+M_x\cos(2\theta_x)$ with $M_x=\frac12q_x^2(c_{x,j}-c_{x,k})\ge0$.
The first two are functions of $r_x$ alone: they multiply $\rho$ by one-site factors, which do not change
mixed derivatives of $\log\rho$, and they do not involve $\tau$. The planar system now has the edge energies
\eqref{eq:Txy} and the site terms $M_x\cos(2\theta_x)$, covered by Lemma~\ref{lem:ginibre}. In the proof of
Lemma~\ref{lem:amplitude}, every coefficient of the planar system ($L_e$ and $M_x$) is nonincreasing in the
relevant amplitude and $M_x$ has no mixed derivative, so all terms remain nonnegative. Pinning, the
crossing identity and the separation of annuli are unaffected by one-site terms. In Lemma~\ref{lem:rn}
the Hamiltonian changes by at most $\beta\delta'(|E_N|+|V_N|)\le62\beta\delta'N^2$, since
$|\sum_ic_{x,i}(\sigma_x^i)^2|\le\max_i|c_{x,i}|\le\beta\delta'$ and $|V_N|\le(4N+1)^2\le21N^2$; so the argument
of Proposition~\ref{prop:connection} runs with
\[
\delta'(\beta)=\min\Bigl\{\frac12,\ \frac{4^{-26}}{124\beta N^2}\Bigr\}\le\delta(\beta),
\]
for the same $N=N(\beta)$, and gives the same $A_0,m_0$. For $\tilde H_\varepsilon$ of~\cite[Definition~1.8(4)]{GarbanMontgolfier2026},
with weight $\exp(-\beta'\tilde H_\varepsilon)$ at $\beta'\le\beta$, we have $\Jv=(1,1,1)$ and
$c_{x,3}=-\beta'\varepsilon$, $c_{x,1}=c_{x,2}=0$, which is aligned for either sign of $\varepsilon$, with
$|c_{x,3}|\le\beta\delta'$ when $|\varepsilon|\le\delta'(\beta)$.
\end{proof}

\begin{remark}\label{rem:align}
The proof of part~(a) for a fixed component $i$ uses only that, for the complementary pair $(j,k)$,
$J^j-J^k$ and all $c_{x,j}-c_{x,k}$ have one weak sign; asking this for all three $i$ is equivalent to
alignment. For example, if $\lambda<1$ is combined with an easy-axis single-ion term, the bound for
$\langle\sigma_x^3\sigma_y^3\rangle$ still holds, but for $i=1,2$ the terms $\cos(\theta_x+\theta_y)$ and $\cos(2\theta_x)$
enter with opposite signs. Then the covariance inequality of Lemma~\ref{lem:ginibre} fails, and the
amplitude law itself need not be log-supermodular: on a path with three sites, with $\lambda=0.98$,
$c_{x,3}=0.08$ at every site, $b_e=4$ and $r=(0.8,0.9,0.5)$, numerical quadrature gives
$\partial_{r_1}\partial_{r_3}\log\rho\approx-1.2\times10^{-5}$, while the aligned control ($c_{x,3}=-0.08$) gives
$\approx+3.9\times10^{-4}$. So this route does not extend to such cases.
\end{remark}

\begin{lemma}[conditional projection]\label{lem:gauss}
Let $2\le k<n$ and write $\sigma\in S^{n-1}$ as $\sigma=(\rho u,v)$ with $u\in S^{k-1}$, $v\in\R^{n-k}$ and
$\rho=(1-|v|^2)^{1/2}$. Under the uniform measure on $S^{n-1}$, $\rho>0$ almost surely, and $u$ is uniform on
$S^{k-1}$ and independent of $v$.
\end{lemma}
\begin{proof}
This is the block form~\cite[Lemma~4.8]{OpenAI215} of~\cite[Proposition~2.4]{AruGarbanSepulveda2025}. Let
$U\in\R^k$ and $W\in\R^{n-k}$ be independent standard Gaussian vectors; $(U,W)/|(U,W)|$ is uniform on $S^{n-1}$.
By rotation invariance $U/|U|$ is uniform on $S^{k-1}$ and independent of $(|U|,W)$, while $v=W/|(U,W)|$ and
$\rho=|U|/|(U,W)|$ are functions of $(|U|,W)$.
\end{proof}

\begin{proof}[Proof of Theorem~\ref{thm:ext}(b)]
Permute coordinates so that the component of interest is the first, and apply Lemma~\ref{lem:gauss} with
$k=3$ at each site. The interaction is $\sum_eb_e[\sum_{i\le3}J^i\rho_x\rho_yu_x^iu_y^i+\sum_{i>3}J^iv_x^iv_y^i]$; given
$v$ the second sum is constant, so the law of $u$ is the three-component model with anisotropy
$(J^1,J^2,J^3)$ and strengths $b_e\rho_x\rho_y\in[0,\beta]$. Theorem~\ref{thm:main} gives
$0\le\langle u_x^1u_y^1\rangle_v\le A_0e^{-m_0\|x-y\|_2}$, and
$\langle\sigma_x^1\sigma_y^1\rangle=\E[\rho_x\rho_y\langle u_x^1u_y^1\rangle_v]$.
\end{proof}

\begin{remark}\label{rem:uniform}
At $\Jv=(1,\dots,1)$ the isotropic law is invariant under permutations of coordinates, so
Theorem~\ref{thm:ext}(b) gives $0\le\langle\sigma_x\cdot\sigma_y\rangle^{(n)}_{G,b}=n\langle\sigma_x^1\sigma_y^1\rangle\le nA_0e^{-m_0\|x-y\|_2}$
for every $n\ge3$, finite $G$ and $b\in[0,\beta]^E$, with $m_0=m_0(\beta)=(\log2)/(25\sqrt2N(\beta))$. This is the
bound~\eqref{eq:polyakov} with $A=nA_0$ and $m=m_0$, obtained from~\cite[Theorem~5.2]{DHY2026} through
Proposition~\ref{prop:dhy}.
\end{remark}

\section{The corollaries}\label{sec:cor}

\begin{proof}[Proof of Corollary~\ref{cor:lambda}]
The $\lambda$-model at $\beta'\le\beta$ has $b\equiv\beta'\in[0,\beta]$ and $\Jv=(1,1,\lambda)$, so Theorem~\ref{thm:main}
applies on every free box when $|1-\lambda|\le\delta(\beta)$. The bound is uniform in the box, so it holds for the
limit superior along boxes and, since $\sigma\mapsto\sigma_0^i\sigma_x^i$ is a bounded continuous local function, for every
subsequential local weak limit. If $\bar\beta(\lambda)$ is any threshold as
in~\cite[Theorem~1.2]{GarbanMontgolfier2026}, which concerns free or Dirichlet limits, the free-boundary
part of its conclusion is incompatible with exponential decay, so $\bar\beta(\lambda)>\beta$ whenever
$0<1-\lambda\le\delta(\beta)$. For $\lambda>1$, exponential decay of $\langle\sigma_x^3\sigma_y^3\rangle$ in free-boundary
limits excludes long-range order there.
\end{proof}

\begin{proof}[Proof of Corollary~\ref{cor:mono}]
By~\cite[Proposition~A.1]{GarbanMontgolfier2026}, with free boundary conditions,
$\lambda\mapsto\langle\sigma_x^1\sigma_y^1\rangle_{\beta,\lambda}$ is nonincreasing and
$\lambda\mapsto\langle\sigma_x^3\sigma_y^3\rangle_{\beta,\lambda}$ is nondecreasing on $[0,\infty)$; this rests on Dunlop's
inequality~\cite[Theorem~1]{Dunlop1976}. At $\lambda=1$, $\langle\sigma_x^i\sigma_y^i\rangle=\frac13\langle\sigma_x\cdot\sigma_y\rangle$,
which~\eqref{eq:polyakov} bounds with $b\equiv\beta$; by Remark~\ref{rem:uniform} one may take $A=3A_0$,
$m=m_0$.
\end{proof}

\section{The physical rate for \texorpdfstring{$n\ge4$}{n>=4}: proof of Theorem~\ref{thm:phys}}\label{sec:phys}

This section adapts the proof of decay for charged observables of the isotropic $O(4)$ model
in~\cite[\S\S3.9--3.10]{OpenAIO4} to anisotropic couplings. The mechanism there is kept: coarse orientation
variables, a test for bad coarse vertices, a disagreement exploration following van den Berg and
Maes~\cite{vdBergMaes1994}, and averaging over a symmetry. New here are the use of the isotropic bad-site
estimate only as an event bound at one scale (Lemmas~\ref{lem:celltransfer}--\ref{lem:badsite}), the
replacement of the averaging over common left rotations by $S^3$ with averaging over $\{\pm1\}$
(Lemmas~\ref{lem:pm1}--\ref{lem:odd}), and the bookkeeping of the threshold (Section~\ref{sec:threshold}).
Throughout, $\omega_n$ is the uniform probability measure on $S^{n-1}$.

\subsection{Systems and the result for \texorpdfstring{$n=4$}{n=4}}

A \emph{system} $\Sys=(V,E,P,\bar\sigma,M)$ consists of a finite $V\subset\Z^2$, a set $E$ of nearest-neighbour edges
with both endpoints in $V$, each with a fixed orientation $e=(x\to y)$, a set $P\subset V$ of pinned sites with
values $\bar\sigma\in(S^{n-1})^P$, and real $n\times n$ matrices $M_e$, $e\in E$. Its Gibbs measure is the probability
measure on $(S^{n-1})^{V\setminus P}$ given by
\[
\langle\cdot\rangle_\Sys\propto\exp\Bigl(\sum_{e=(x\to y)\in E}\sigma_x^{\mathsf T}M_e\sigma_y\Bigr)\prod_{z\in V\setminus P}\omega_n(d\sigma_z),
\qquad\sigma_p:=\bar\sigma_p\ (p\in P).
\]
The system is \emph{$(\beta,\theta)$-admissible} if for every $e$ there is $b_e\in[-\beta,\beta]$ with
$\|M_e-b_eI\|_{\mathrm{op}}\le\theta$. A \emph{torus system} is defined in the same way on
$(\Z/L_1\Z)\times(\Z/L_2\Z)$ with all nearest-neighbour edges and $P=\emptyset$; $d_T$ is the cyclic
maximum-norm distance. An observable on $A$ is a bounded measurable function $O$ of $\sigma_A$; it is
\emph{odd} if $O(-\sigma_A)=-O(\sigma_A)$. The systems of Theorem~\ref{thm:phys} have
$M_e=\mathrm{diag}(m_e^1,\dots,m_e^n)$ with $|m_e^k-b_e|\le\beta\delta$, so they are $(\beta,\beta\delta)$-admissible.

\begin{theorem}\label{thm:main4}
Assume (I1)--(I3) of Section~\ref{sec:o4inputs}. There are absolute constants $c_1>0$ and $A_*,C_\Xi<\infty$
such that, with $\Xi(\beta)$ as in Theorem~\ref{thm:phys}, the following hold whenever $\beta>0$, $\theta\ge0$ and
$\theta\,\Xi(\beta)^2\le c_1$. For every $(\beta,\theta)$-admissible system with $n=4$, every odd observable $O$ and
every observable $O'$ on $S^3$:
\begin{enumerate}
\item[(a)] $|\langle O(\sigma_x)\rangle|\le A_*\|O\|_\infty\exp(-\dist_\infty(x,P)/\Xi(\beta))$ for $x\in V\setminus P$; the
left side vanishes if $P=\emptyset$;
\item[(b)] for $x,y\in V\setminus P$,
\[
|\langle O(\sigma_x)O'(\sigma_y)\rangle|\le A_*\|O\|_\infty\|O'\|_\infty\exp\bigl(-\min\{\|x-y\|_\infty,\dist_\infty(x,P)\}/\Xi(\beta)\bigr);
\]
\item[(c)] on a torus system with $(\beta,\theta)$-admissible couplings, the bound of (b) holds with
$\min\{\|x-y\|_\infty,\dist_\infty(x,P)\}$ replaced by
\[
r_T(x,y):=\min\Bigl\{d_T(x,y),\Bigl\lfloor\frac{\min\{L_1,L_2\}-1}{2}\Bigr\rfloor\Bigr\}.
\]
\end{enumerate}
\end{theorem}

\subsection{The coarse structure of the \texorpdfstring{$O(4)$}{O(4)} proof and the inputs}\label{sec:o4inputs}

Identify $\R^4$ with the quaternions $\HH$ by sending the standard basis to $(1,\mathrm i,\mathrm j,\mathrm k)$.
Then $S^3$ is the group of unit quaternions, $\omega_4$ is its Haar measure,
$\sigma\cdot\sigma'=\mathrm{Re}(\sigma'\bar\sigma)$ for unit $\sigma,\sigma'$, and left multiplication by the central
element $-1$ is $\sigma\mapsto-\sigma$.

Fix $n=4$, $\beta>0$, $\theta\ge0$ and a $(\beta,\theta)$-admissible system on a \emph{pinned square}: $V=Q(S)=\{0,\dots,S\}^2$,
$E$ all nearest-neighbour edges of $Q(S)$, $P=\partial Q(S)$ (the sites with a coordinate in $\{0,S\}$) and
arbitrary $\bar\sigma$; general systems are reduced to this case in Section~\ref{sec:general}. Its \emph{reference
isotropic system} has the same $(V,E,P,\bar\sigma)$ and $M_e$ replaced by $b_eI$, that is, edge energies
$b_e\sigma_x\cdot\sigma_y=b_e\mathrm{Re}(\sigma_y\bar\sigma_x)$. This is the class of~\cite[\S3.1]{OpenAIO4} with untwisted bonds
($U_e=1$) and signed coefficients $|b_e|\le b:=\beta$. The arguments of~\cite[\S3]{OpenAIO4} depend on fixed
choices (exponents $D,C_0$, and the collars and profile families of~\cite[\S\S3.9--3.10]{OpenAIO4}), and its
constants depend only on them, not on the lattice size, the couplings or the pins~\cite[p.~13]{OpenAIO4}. We
keep these choices. For $B$ above the threshold of~\cite[\S3]{OpenAIO4} we use:

\begin{itemize}
\item[(I1)] \cite[(3.54)]{OpenAIO4}: for $|b_e|\le b\le B$ there is an integer $R_*$, depending only on $b$,
$B$ and the fixed choices, such that pinned squares of side $R_*$ in the class have constant-field stiffness
$H_{R_*}\le B^{-D}I$ and $\log R_*\le\pi b+C_D\sqrt{B\log B}$, where $C_D$ depends only on the fixed choices.
The coarse scale is $s:=R_*M_*$ with $M_*=\lceil B^{D+C_0}\rceil$~\cite[\S3.9]{OpenAIO4}. Only the bound on
$\log R_*$ is used directly below (Lemma~\ref{lem:scale}); the stiffness bound enters through (I3).
\end{itemize}

\emph{Fixed choices.} Let $\varepsilon_*:=1/32$ and $q:=1-e^{-\varepsilon_*}\le1/32$. For an interior coarse vertex $X$
(below), $\coll(X)$ is the fixed collar of~\cite[\S3.9]{OpenAIO4}: a set of at most $K_{\mathrm{coll}}$ coarse
cells containing the four cells at $X$ (the \emph{star} of $X$), each corner of which is at grid distance at
most $r_{\mathrm{coll}}$ from $X$. We read it as a block of cells without holes, so that the profiles of (I2),
which are the identity on and outside the outer collar boundary, are the identity outside the set $I(X)$
defined below; otherwise replace $\coll(X)$ by the set of cells whose rings meet the support of those
profiles, and nothing below changes. Put $C_{\mathrm{coll}}:=(2r_{\mathrm{coll}}+1)^2$ and $w:=r_{\mathrm{coll}}+2$.
Choose $B_1\ge e$ at least the threshold of~\cite{OpenAIO4} and so large that
$\zeta_{1/128}(B)\le\frac12\cdot32^{-C_{\mathrm{coll}}}$ for all $B\ge B_1$, which (I3) below allows. Put
$B:=\max\{\beta,B_1\}$ and let $s$ be the scale of (I1) with $b=\beta$.

\emph{Tiling and coarse graph.} If $S\ge s$, split $[0,S]$ into consecutive integer intervals
$0=x_0<\dots<x_k=S$ with lengths in $[s,2s]$. The \emph{coarse vertices} are the points $(x_i,x_j)$, and the
\emph{cells} are the rectangles $[x_i,x_{i+1}]\times[x_j,x_{j+1}]\cap\Z^2$. A cell $c$ has \emph{ring} $\partial c$ (its
boundary sites) and \emph{interior} $c^\circ$. A coarse vertex is \emph{pinned} if $i$ or $j$ lies in $\{0,k\}$, and
\emph{free} otherwise; $d_G$ is the grid maximum-norm distance between coarse vertices. Let $\mathcal G$ be the
graph on the free coarse vertices with $d_G$-adjacency (at most $8$ neighbours); the free corners of a cell
form a clique of $\mathcal G$. The ring sites are $R=\bigcup_c\partial c$; the pinned ones form $\partial Q(S)$ and
the others form $R_f$. Let $W:=\{X\text{ free}:d_G(X,\text{pinned})\le w\}$ (the \emph{seed layer}) and
$U:=\{X\text{ free}:d_G(X,\text{pinned})>w\}$ (the \emph{tested} vertices). For $X\in U$ let $I(X)$ be the set of
ring sites that lie on $\partial c$ for some $c\in\coll(X)$ and on no $\partial c'$ with $c'\notin\coll(X)$. Then:
\begin{itemize}
\item[(F1)] for $X\in U$ every cell of $\coll(X)$ has four free corners (at $d_G$-distance at least $3$ from the
pinned vertices), and $I(X)\cap\partial Q(S)=\emptyset$;
\item[(F2)] every cell with a corner in $U$ has four free corners;
\item[(F3)] if $d_G(X,Y)>2r_{\mathrm{coll}}$ then $\coll(X)\cap\coll(Y)=\emptyset$; distinct coarse vertices whose indices
agree modulo $2r_{\mathrm{coll}}+1$ are at $d_G$-distance at least $2r_{\mathrm{coll}}+1$, so $U$ splits into
$C_{\mathrm{coll}}$ classes, each with pairwise disjoint collars.
\end{itemize}
The width $w$ of the seed layer is chosen so that tested collars contain no pinned site, as~\cite[\S3.9]{OpenAIO4}
requires.

\emph{Cell actions}~\cite[\S\S3.1, 3.3]{OpenAIO4}. For an edge $e$ with both endpoints in a cell $c$ let
$\kappa_{c,e}=\frac12$ if both endpoints lie on one side of $\partial c$ and $\kappa_{c,e}=1$ otherwise. Every edge
that does not join two pinned sites has $\sum_c\kappa_{c,e}=1$. For $\bullet\in\{\an,\iso\}$ (anisotropic and
isotropic) let
\begin{gather*}
A^{\an}_c(\sigma_c)=\sum_{e\subset c}\kappa_{c,e}\,\sigma_x^{\mathsf T}M_e\sigma_y,\qquad A^{\iso}_c(\sigma_c)=\sum_{e\subset c}\kappa_{c,e}\,b_e\,\sigma_x\cdot\sigma_y,\\
F^\bullet_c(\varrho)=-\log\int e^{A^\bullet_c(\varrho,\iota)}\prod_{z\in c^\circ}\omega_4(d\iota_z),
\end{gather*}
$F^\bullet=\sum_cF^\bullet_c$, and let $\nu^\bullet(d\omega)\propto e^{-F^\bullet(\omega)}\prod_{z\in R_f}\omega_4(d\omega_z)$ be the ring law,
with $\omega=\bar\sigma$ on $\partial Q(S)$. Given the rings, the cell interiors are independent with laws
$\mu^\bullet_c(d\iota\mid\varrho)\propto e^{A^\bullet_c(\varrho,\iota)}d\iota$, since a site of $c^\circ$ has all its neighbours in $c$.

\emph{Coarse orientations}~\cite[\S3.10]{OpenAIO4}. Let $g\in(S^3)^{\{\text{free coarse vertices}\}}$, with
$g_X:=1$ at pinned $X$. On a coarse edge $X\to Y$ (a side of a cell, positively oriented) put
$\Phi(g;t)=g_X\exp[\eta(t)\ell(g_X^{-1}g_Y)]$, $t\in[0,1]$, where $\eta$ is a fixed smooth function, constant near
the endpoints, with $\eta(0)=0$ and $\eta(1)=1$, and $\ell$ is a fixed measurable logarithm of norm at most $\pi$
with $\ell(1)=0$. Write $\omega=\Phi(g)\tilde\omega$ for the sitewise left multiplication $\omega_z=\Phi(g;t_z)\tilde\omega_z$ of
the free ring sites, $t_z$ being the position of $z$ on its coarse edge. For fixed $g$ this preserves Haar
measure, and $\Phi(g)\tilde\omega|_{\partial c}$ depends on $g$ only through the corners of $c$. Let the pair
$(g,\tilde\omega)$ have density proportional to $e^{-F^{\an}(\Phi(g)\tilde\omega)}$ with respect to Haar measure. The change of
variables $(g,\tilde\omega)\mapsto(g,\Phi(g)\tilde\omega)$ has unit Jacobian, so $(g,\omega)$ has density proportional to
$e^{-F^{\an}(\omega)}$: \emph{$g$ is Haar distributed and independent of $\omega$, and $\omega\sim\nu^{\an}$}. Let
$\nu_{\tilde\omega}$ be the conditional law of $g$ given $\tilde\omega$; its density is proportional to
$\prod_c\psi_c(g)$ with $\psi_c(g)=\exp(-F^{\an}_c(\Phi(g)\tilde\omega|_{\partial c}))$.

\emph{Bad vertices}~\cite[\S3.10]{OpenAIO4}, with the anisotropic action. A vertex $X\in U$ is \emph{bad} for
$\tilde\omega$ if there are assignments $g',g''$ differing only at $X$, and logarithms of norm at most $\pi$ satisfying
the endpoint equations on the edges at $X$, with the same data on the other edges, such that the
interpolations $\Phi',\Phi''$ they define satisfy
$\bigl|\sum_{c\in\mathrm{star}(X)}[F^{\an}_c(\Phi'(g')\tilde\omega)-F^{\an}_c(\Phi''(g'')\tilde\omega)]\bigr|>\varepsilon_*$; otherwise $X$
is \emph{good}. At a good vertex, with the other orientations fixed, $\sum_{c\ni X}F^{\an}_c(\Phi(g)\tilde\omega)$ varies by at
most $\varepsilon_*$ as $g_X$ ranges over $S^3$. The set of valid parameters is compact and the star action is
continuous in them and in $\tilde\omega$, so $\{X\text{ bad}\}$ is open, hence measurable.

\begin{itemize}
\item[(I2)] \cite[\S3.10.1]{OpenAIO4}: for every $X\in U$ there are a compact parameter set $\Lambda_X$, a finite
union of coordinate rectangles, and ring rotations $\phi_\lambda$, $\lambda\in\Lambda_X$ (sitewise left multiplications,
smooth in $\lambda$, equal to the identity outside $I(X)$), such that for every $g$, all $g',g''$ differing only at
$X$ and all admissible logarithms there are $\lambda',\lambda''\in\Lambda_X$ (both extensions are built in one
chart; $\Lambda_X$ may carry a label $j\in\{1,2\}$ for the two trials) with $\phi_{\lambda'}=\phi_{\lambda''}$ on $\partial c$ for every
$c\notin\mathrm{star}(X)$, and $\phi_{\lambda'}=\Phi'(g')\Phi(g)^{-1}$, $\phi_{\lambda''}=\Phi''(g'')\Phi(g)^{-1}$ on the rings
of the star. This is a deterministic construction; its constants do not depend on the lattice size, the
couplings or the configurations.
\item[(I3)] \cite[(3.58) and pp.~26--28]{OpenAIO4}: for the reference isotropic system, every $X\in U$ and every
$\varepsilon>0$, with $R_X:=R_f\setminus I(X)$,
\[
\nu^{\iso}\Bigl(\sup_{\lambda\in\Lambda_X}\bigl|F^{\iso}(\phi_\lambda\omega)-F^{\iso}(\omega)\bigr|>\varepsilon\ \Big|\ \omega_{R_X}\Bigr)\le\zeta_\varepsilon(B)
\quad\text{for every configuration }\omega_{R_X},
\]
where $\zeta_\varepsilon(B)\to0$ as $B\to\infty$ and $\zeta_\varepsilon(B)$ depends only on $\varepsilon$, $B$ and the fixed choices
(this includes the fixed factor from the finitely many parameter rectangles~\cite[p.~28]{OpenAIO4}).
\end{itemize}
In~\cite{OpenAIO4}, (3.58) is stated for a collar without pins, uniformly under conditioning on all rings
outside the collar, and is then applied to the whole bad-site test~\cite[p.~28]{OpenAIO4}; by (F1) tested
collars have no pins. Given $\omega_{R_X}$, the conditional law of $\omega_{I(X)}$ involves only the cells of $\coll(X)$,
so the estimate does not depend on the ambient domain. In~\cite{OpenAIO4} the proof of (3.58) uses the
stiffness bound in (3.54) and~\cite[(3.10), (3.11), (3.13), (3.14)]{OpenAIO4}, that is, the iteration
of~\cite[\S\S3.1--3.8]{OpenAIO4}; we assume (I3) as stated. Apart from the group structure of $S^3$ used for
the coarse orientations, (I1) and (I3) are the only places where the dimension two and the value $n=4$ enter.

\subsection{Transfer of the bad-site estimate}

\begin{lemma}[cell transfer]\label{lem:celltransfer}
For every cell $c$ with sides $a_c,b_c\le2s$ and every ring configuration $\varrho$,
$|F^{\an}_c(\varrho)-F^{\iso}_c(\varrho)|\le2a_cb_c\theta\le8\theta s^2$. For $X\in U$ and every sitewise left multiplication $\phi$ of
the ring that is the identity outside $I(X)$,
\[
\bigl|[F^{\an}(\phi\omega)-F^{\an}(\omega)]-[F^{\iso}(\phi\omega)-F^{\iso}(\omega)]\bigr|\le16K_{\mathrm{coll}}\theta s^2.
\]
\end{lemma}
\begin{proof}
For unit vectors $|\sigma^{\mathsf T}(M_e-b_eI)\sigma'|\le\theta$, so $|A^{\an}_c-A^{\iso}_c|\le\theta\sum_{e\subset c}\kappa_{c,e}$. A rectangle with
sides $a,b$ has $a(b+1)+b(a+1)$ edges, of which $2(a+b)$ lie on its boundary and have weight $\frac12$, so
$\sum_e\kappa_{c,e}=2ab\le8s^2$. Since $e^{-\|\Delta\|_\infty}\int e^{A^{\iso}_c}\le\int e^{A^{\an}_c}\le e^{\|\Delta\|_\infty}\int e^{A^{\iso}_c}$ with
$\Delta=A^{\an}_c-A^{\iso}_c$, the first claim follows. For the second,
$F^\bullet(\phi\omega)-F^\bullet(\omega)=\sum_c[F^\bullet_c(\phi\omega)-F^\bullet_c(\omega)]$ over the cells whose rings meet $I(X)$; these
lie in $\coll(X)$, so there are at most $K_{\mathrm{coll}}$ of them, and each contributes at most $2\cdot8\theta s^2$.
\end{proof}

\begin{lemma}[bad-site bound]\label{lem:badsite}
Assume $16K_{\mathrm{coll}}\theta s^2\le\varepsilon_*/4$. There are events $G_X$, $X\in U$, of the ring variables such that:
(a) if $X$ is bad for $\tilde\omega$, then $\Phi(g)\tilde\omega\in G_X$ for every $g$; (b) $G_X$ depends only on $\omega$
restricted to $\bigcup_{c\in\coll(X)}\partial c$; (c) for every finite $\mathcal S\subset U$,
\[
\nu^{\an}\Bigl(\bigcap_{X\in\mathcal S}G_X\Bigr)\le\zeta_{\an}^{\lceil|\mathcal S|/C_{\mathrm{coll}}\rceil},\qquad
\zeta_{\an}:=e^{\varepsilon_*/4}\zeta_{\varepsilon_*/4}(B)\le32^{-C_{\mathrm{coll}}}.
\]
\end{lemma}
\begin{proof}
\emph{Step 1 (deterministic).} Let $g$ be arbitrary, $\omega=\Phi(g)\tilde\omega$, and let $g',g''$ and the logarithms
be as in the definition of a bad vertex. With $\lambda',\lambda''$ from (I2), for every cell-additive $F=\sum_cF_c$,
\[
F(\phi_{\lambda'}\omega)-F(\phi_{\lambda''}\omega)=\sum_{c\in\mathrm{star}(X)}\bigl[F_c(\Phi'(g')\tilde\omega)-F_c(\Phi''(g'')\tilde\omega)\bigr],
\]
since the two rotations agree on the rings of the cells outside the star, as~\cite[p.~28]{OpenAIO4} notes. Let
$G_X:=\{\omega:\sup_{\lambda\in\Lambda_X}|F^{\an}(\phi_\lambda\omega)-F^{\an}(\omega)|>\varepsilon_*/2\}$. If $X$ is bad, the left side exceeds $\varepsilon_*$ for
some choice, so one of the two displacements exceeds $\varepsilon_*/2$; this is~(a). Since $\phi_\lambda$ is the identity
outside $I(X)$, only cells of $\coll(X)$ change, which is~(b). $G_X$ is open, since $\Lambda_X$ is a finite union
of compact rectangles and the displacement is jointly continuous.

\emph{Step 2 (transfer).} By Lemma~\ref{lem:celltransfer},
$G_X\subset G_X^{\iso}:=\{\omega:\sup_{\lambda\in\Lambda_X}|F^{\iso}(\phi_\lambda\omega)-F^{\iso}(\omega)|>\varepsilon_*/4\}$.

\emph{Step 3 (isotropic bound).} By (I3) with $\varepsilon=\varepsilon_*/4$, $\nu^{\iso}(G_X^{\iso}\mid\omega_{R_X})\le\zeta_{\varepsilon_*/4}(B)$ for every
$\omega_{R_X}$, where $R_X=R_f\setminus I(X)$.

\emph{Step 4 (conditional density).} Fix $\omega_{R_X}$. For $\bullet\in\{\an,\iso\}$ the conditional ring law is
$\nu^\bullet(d\omega_I\mid\omega_{R_X})\propto\exp(-\sum_{c\in\coll(X)}F^\bullet_c(\omega))\prod_{z\in I}\omega_4(d\omega_z)$, $I=I(X)$, since the other
cells do not involve $\omega_I$. By Lemma~\ref{lem:celltransfer} the Radon--Nikodym derivative of
$\nu^{\an}(\cdot\mid\omega_{R_X})$ with respect to $\nu^{\iso}(\cdot\mid\omega_{R_X})$ is at most
$\exp(2\sum_{c\in\coll(X)}\|F^{\an}_c-F^{\iso}_c\|_\infty)\le e^{16K_{\mathrm{coll}}\theta s^2}\le e^{\varepsilon_*/4}$. Hence
$\nu^{\an}(G_X\mid\omega_{R_X})\le e^{\varepsilon_*/4}\nu^{\iso}(G_X^{\iso}\mid\omega_{R_X})\le\zeta_{\an}$. The anisotropic law of $\omega_{R_X}$ never enters.

\emph{Step 5 (iterated conditioning).} By (F3), choose $\mathcal S'\subset\mathcal S$ with
$|\mathcal S'|\ge\lceil|\mathcal S|/C_{\mathrm{coll}}\rceil$ and pairwise disjoint collars, and enumerate it as $X_1,\dots,X_m$.
For $j<m$, $G_{X_j}$ depends on the rings of $\coll(X_j)$, which avoid $I(X_m)$, because a site of $I(X_m)$ lies
only on rings of cells of $\coll(X_m)$. So
$\nu^{\an}(\bigcap_{j\le m}G_{X_j})=\E^{\an}[\prod_{j<m}\1_{G_{X_j}}\,\nu^{\an}(G_{X_m}\mid\omega_{R_{X_m}})]\le\zeta_{\an}\nu^{\an}(\bigcap_{j<m}G_{X_j})$, and
induction gives $\zeta_{\an}^m$. Finally $\zeta_{\an}\le e^{1/128}\cdot\frac12\cdot32^{-C_{\mathrm{coll}}}\le32^{-C_{\mathrm{coll}}}$.
\end{proof}

\subsection{Odd observables and the coupling}

\begin{lemma}[the symmetry $\sigma\mapsto-\sigma$]\label{lem:pm1}
(a) $A^{\an}_c(-\varrho,-\iota)=A^{\an}_c(\varrho,\iota)$; hence $F^{\an}_c(-\varrho)=F^{\an}_c(\varrho)$, and $\mu^{\an}_c(\cdot\mid-\varrho)$ is the image of
$\mu^{\an}_c(\cdot\mid\varrho)$ under $\iota\mapsto-\iota$. (b) Let $T$ negate $g$ at every free coarse vertex. On every coarse edge
with both endpoints free, $\Phi(Tg;t)=-\Phi(g;t)$; hence for every cell $c$ with four free corners
$\Phi(Tg)\tilde\omega|_{\partial c}=-\Phi(g)\tilde\omega|_{\partial c}$ and $\psi_c(Tg)=\psi_c(g)$. (c) The law $\nu_{-1}:=T_\#\nu_{\tilde\omega}$ has
density proportional to $\prod_c\psi_c(Tg)$, whose factor for every cell with four free corners, in particular
by (F2) for every cell with a corner in $U$, equals that of $\nu_{\tilde\omega}$.
\end{lemma}
\begin{proof}
(a) Each term $(-\sigma)^{\mathsf T}M_e(-\sigma')=\sigma^{\mathsf T}M_e\sigma'$ is bilinear, and $\omega_4$ is invariant under
$\sigma\mapsto-\sigma$. (b) Since $-1$ is central, $(-g_X)^{-1}(-g_Y)=g_X^{-1}g_Y$, so
$\Phi(Tg;t)=(-g_X)\exp[\eta(t)\ell(g_X^{-1}g_Y)]=-\Phi(g;t)$; then use (a). (c) $T$ is a Haar-preserving
involution.
\end{proof}

This is the only place where the structure of the anisotropy is used beyond Lemma~\ref{lem:celltransfer}: the
interaction is even. The common left multiplication by a general $k\in S^3$, which~\cite{OpenAIO4} averages
over, does not preserve $\sigma^{\mathsf T}M_e\sigma'$ unless $M_e$ commutes with left multiplication by $k$.

\begin{lemma}[odd observables]\label{lem:odd}
Let $O$ be an odd observable on $A\subset Q(S)\setminus\partial Q(S)$, let $C_A$ be the set of cells whose rectangle meets
$A$, and let $A^*$ be the set of corners of cells in $C_A$, so $|A^*|\le16|A|$. Put
$O^\sharp(\omega):=\int O(\sigma_A)\prod_{c\in C_A}\mu^{\an}_c(d\iota_c\mid\omega|_{\partial c})$, where $\sigma_A$ is read off $(\omega,\iota)$, and
$O^*(g):=O^\sharp(\Phi(g)\tilde\omega)$. Then (a) $\langle O\rangle=\E_{\tilde\omega}\bigl[\E_{\nu_{\tilde\omega}}O^*\bigr]$, where $\tilde\omega$ has its
marginal law under the joint law; (b) $|O^*|\le\|O\|_\infty$ and $O^*$ depends on $g$ only through $g_{A^*}$; (c) if
every cell of $C_A$ has four free corners, then $O^*(Tg)=-O^*(g)$.
\end{lemma}
\begin{proof}
(a) The Gibbs law is the ring law $\nu^{\an}$ followed by the independent interior kernels, so
$\langle O\rangle=\E_{\nu^{\an}}[O^\sharp]$, and $\Phi(g)\tilde\omega\sim\nu^{\an}$ under the joint law. (b) is clear. (c) By
Lemma~\ref{lem:pm1}(a) and the substitution $\iota\mapsto-\iota$, $O^\sharp(-\omega)=-O^\sharp(\omega)$; by Lemma~\ref{lem:pm1}(b),
$\Phi(Tg)\tilde\omega=-\Phi(g)\tilde\omega$ on the rings of the cells of $C_A$.
\end{proof}

\begin{lemma}[disagreement coupling]\label{lem:coupling}
Let $\mathcal G$ be a finite graph with vertex set $U\sqcup W$, let $(Y,\varpi)$ be a compact probability space, and let
$\mathcal C$ be a finite family of cliques of $\mathcal G$. For $j=1,2$ let
$\mu^j(dg)=Z_j^{-1}\prod_{c\in\mathcal C}\psi^j_c(g_c)\,\varpi^{\otimes(U\sqcup W)}(dg)$, with each $\psi^j_c$ measurable and bounded above
and below by positive constants, and with $\psi^1_c=\psi^2_c$ whenever $c\cap U\ne\emptyset$. Let $p\in[0,1]^U$ satisfy
$\mu^j(g_v\in\cdot\mid g_{\mathrm{rest}})\ge(1-p_v)\varpi$ for every $v\in U$, every $j$ and every configuration of the rest. Then for
every $A'\subset U$ and every bounded measurable $H$ of $g_{A'}$,
\[
|\mu^1(H)-\mu^2(H)|\le2\|H\|_\infty\,\Pp_p(A'\lr W),
\]
where under $\Pp_p$ the vertices of $U$ are independently open with probabilities $p_v$, and $\{A'\lr W\}$ is the
event that there are distinct open $v_0,\dots,v_{m-1}\in U$, $m\ge1$, with $v_0\in A'$, $v_{i-1}$ adjacent to $v_i$, and
$v_{m-1}$ adjacent to $W$.
\end{lemma}
\begin{proof}
The mechanism is that of van den Berg and Maes~\cite[Theorem~1]{vdBergMaes1994}, which is stated for finite
single-site spaces; the continuous version with adaptive exploration is~\cite[\S3.10.2]{OpenAIO4}. We give
the argument. Let $(U_v)_{v\in U}$ be independent and uniform on $[0,1]$. Draw $X^1_W$ and $X^2_W$ independently from
the $W$-marginals of $\mu^1$ and $\mu^2$, and declare $W$ revealed and marked. While some unrevealed vertex of $U$
is adjacent to a marked vertex, let $v$ be the first such vertex in a fixed enumeration and let
$m^j:=\mu^j(g_v\in\cdot\mid g_{\mathrm{revealed}}=X^j_{\mathrm{revealed}})$; as an average of one-site conditionals,
$m^j\ge(1-p_v)\varpi$. If $U_v\ge p_v$, draw $y\sim\varpi$ and set $X^1_v=X^2_v=y$. Otherwise draw $X^j_v$ independently from
$(m^j-(1-p_v)\varpi)/p_v$ and mark $v$. Reveal $v$. When this stops, let $Q$ be the set of unrevealed vertices.
Every neighbour of $Q$ is revealed and unmarked, so $X^1=X^2$ on the boundary of $Q$. Since the factors are
cliques, the conditional law of $g_Q$ given the revealed values is proportional to $\prod_{c\cap Q\ne\emptyset}\psi^j_c$ and
depends only on the values on the boundary of $Q$; these factors meet $U$, so they agree for $j=1,2$, and we
draw $X^1_Q=X^2_Q$ from the common kernel.

\emph{Marginals.} Fix $j$ and a bounded $H$, and let $M_t$ be the $\mu^j$-conditional expectation of $H$ given the
values revealed after $t$ steps. Given the history, the next vertex is determined, $U_v$ is independent, and
$X^j_v$ has law $(1-p_v)\varpi+p_v(m^j-(1-p_v)\varpi)/p_v=m^j$; so $(M_t)$ is a martingale, the final draw has the
conditional law, and $\E H(X^j)=\E M_0=\mu^j(H)$. \emph{Disagreement.} If $X^1_a\ne X^2_a$ for some $a\in A'$, then
$a$ is marked; each marked vertex of $U$ was revealed while adjacent to an earlier marked vertex or to $W$,
and following these predecessors gives distinct marked $v_0=a,v_1,\dots$ ending next to $W$. Marked vertices
have $U_v<p_v$, so $\{X^1_{A'}\ne X^2_{A'}\}$ is contained in an event of $(U_v)$ of probability at most
$\Pp_p(A'\lr W)$.
\end{proof}

\begin{proposition}[pinned decay]\label{prop:pinneddecay}
Assume $16K_{\mathrm{coll}}\theta s^2\le\varepsilon_*/4$. For every odd observable $O$ on $A\subset Q(S)\setminus\partial Q(S)$, with
$d:=\dist_\infty(A,\partial Q(S))$,
\[
|\langle O\rangle|\le2^{3+w}|A|\,\|O\|_\infty\exp\Bigl(-\frac{(\log2)\,d}{2s}\Bigr).
\]
\end{proposition}
\begin{proof}
If $d\le2s(w+2)$, which includes the case $S<s$, then $|\langle O\rangle|\le\|O\|_\infty\le2^{w+2}e^{-(\log2)d/(2s)}\|O\|_\infty$. Let
$d>2s(w+2)$; then $S\ge2d>s$ and the tiling exists.

\emph{Geometry.} A corner $X$ of a cell of $C_A$ is within sup-distance $2s$ of $A$, so
$\dist_\infty(X,\partial Q(S))\ge d-2s$; each grid step moves at most $2s$, so
$d_G(X,\text{pinned})\ge d/(2s)-1>w+1$. Hence $A^*\subset U$ and every cell of $C_A$ has four free corners.

\emph{Oddness.} Fix $\tilde\omega$ and write $\nu=\nu_{\tilde\omega}$. By Lemma~\ref{lem:odd}(c),
$\E_\nu O^*=-\E_\nu[O^*\circ T]=-\E_{\nu_{-1}}O^*$, so $|\E_\nu O^*|=\frac12|\E_\nu O^*-\E_{\nu_{-1}}O^*|$.

\emph{Coupling.} Apply Lemma~\ref{lem:coupling} with $\mathcal G,U,W$ as above, $\mathcal C$ the cells restricted to their
free corners, $\varpi=\omega_4$, $\mu^1=\nu$, $\mu^2=\nu_{-1}$, $\psi^1_c=\psi_c$, $\psi^2_c=\psi_c\circ T$, $A'=A^*$ and $H=O^*$. The factors
agree on every cell meeting $U$ (Lemma~\ref{lem:pm1}(c)) and are bounded above and below. Put $p_v=q$ if $v$
is good and $p_v=1$ if $v$ is bad. At $v\in U$ the one-site conditional density of $\nu$ is proportional to
$\prod_{c\ni v}\psi_c$ as a function of $g_v$; if $v$ is good its exponent varies by at most $\varepsilon_*$, so the density is
at least $e^{-\varepsilon_*}=1-q$, and the same holds for $\nu_{-1}$, whose factors at $v$ coincide with those of $\nu$.
Hence $|\E_\nu O^*|\le\|O\|_\infty\Pp_{p(\tilde\omega)}(A^*\lr W)$.

\emph{Average over $\tilde\omega$.} By Lemma~\ref{lem:odd}(a) and a union bound,
$|\langle O\rangle|\le\|O\|_\infty\sum_\pi\E_{\tilde\omega}[\prod_{v\in\pi}p_v(\tilde\omega)]$, summed over sequences
$\pi=(v_0,\dots,v_{m-1})$ as in Lemma~\ref{lem:coupling}. Here $\prod_{v\in\pi}p_v=\prod_v(q+(1-q)\1_{v\text{ bad}})$, which by
Lemma~\ref{lem:badsite}(a) is at most $\prod_v(q+(1-q)\1_{G_v}(\Phi(g)\tilde\omega))$ for every $g$. Taking $g$ jointly
distributed with $\tilde\omega$, so that $\Phi(g)\tilde\omega\sim\nu^{\an}$, expanding the product and using
Lemma~\ref{lem:badsite}(c) and $\zeta_{\an}\le1$,
\[
\E_{\tilde\omega}\Bigl[\prod_{v\in\pi}p_v\Bigr]\le\sum_{\mathcal S\subset\pi}q^{m-|\mathcal S|}(1-q)^{|\mathcal S|}\nu^{\an}\Bigl(\bigcap_{v\in\mathcal S}G_v\Bigr)
\le\bigl(q+(1-q)\zeta_{\an}^{1/C_{\mathrm{coll}}}\bigr)^m\le16^{-m},
\]
since $q\le1/32$ and $\zeta_{\an}^{1/C_{\mathrm{coll}}}\le1/32$. This is the analogue of~\cite[(3.60)]{OpenAIO4}.

\emph{Path count.} Every vertex of $W$ is within $d_G$-distance $w$ of a pinned vertex, so every such $\pi$
starting at $v_0\in A^*$ has $m\ge d_G(v_0,\text{pinned})-w\ge d/(2s)-1-w=:k_0$. There are at most $8^{m-1}$ such $\pi$
for each start, so the sum is at most
$|A^*|\sum_{m\ge\lceil k_0\rceil}8^{m-1}16^{-m}\le|A^*|2^{-k_0}/4\le4|A|\,2^{1+w}2^{-d/(2s)}$.
\end{proof}

With $\theta=0$ this is the decay of~\cite[\S3.10]{OpenAIO4} restricted to odd observables, with the
seed-layer width fixed as $w$.

\subsection{Scale and threshold}\label{sec:threshold}

\begin{lemma}[scale]\label{lem:scale}
There is an absolute $C_\Xi$ such that $\xi_s:=2s/\log2\le\Xi(\beta)$ for all $\beta>0$. Consequently
$e^{-(\log2)d/(2s)}\le e^{-d/\Xi(\beta)}$, and $\theta\,\Xi(\beta)^2\le c_1:=1/(256K_{\mathrm{coll}})$ implies
$16K_{\mathrm{coll}}\theta s^2\le\varepsilon_*/4$.
\end{lemma}
\begin{proof}
By (I1) and $M_*\le2B^{D+C_0}$, $\log s\le\pi\beta+C_D\sqrt{B\log B}+(D+C_0)\log B+\log2$. If $\beta\ge B_1$, then $B=\beta$,
$\log\beta\le\sqrt{\beta\log\beta}\le\sqrt{(1+\beta)\log(2+\beta)}$, and
$\log s\le\pi\beta+(C_D+D+C_0)\sqrt{(1+\beta)\log(2+\beta)}+\log2$. If $\beta<B_1$, then
$\log s\le\pi\beta+K_1$ with $K_1:=C_D\sqrt{B_1\log B_1}+(D+C_0)\log B_1+\log2$. Since
$\sqrt{(1+\beta)\log(2+\beta)}\ge\sqrt{\log2}$, the choice $C_\Xi:=C_D+D+C_0+(K_1+\log(4/\log2))/\sqrt{\log2}$ gives
$\log\xi_s\le\log\Xi(\beta)$ in both cases. Finally $s^2=(\log2/2)^2\xi_s^2\le\Xi^2/8$, so
$16K_{\mathrm{coll}}\theta s^2\le2K_{\mathrm{coll}}\theta\Xi^2\le1/128$.
\end{proof}

\begin{lemma}[threshold]\label{lem:threshold}
Put $L:=|\log\delta|$; the function $f(\beta):=\beta\,\Xi(\beta)^2$ is increasing.
(a) For every $\varepsilon\in(0,1)$ there is $\delta_0(\varepsilon)\in(0,1)$ such that $0<\delta<\delta_0(\varepsilon)$ and
$0<\beta\le(1-\varepsilon)L/(2\pi)$ imply $\beta\delta\,\Xi(\beta)^2\le c_1$.
(b) With $K:=3C_\Xi+2$ there is $\delta_1>0$ such that $0<\delta<\delta_1$ and
$0<\beta\le(L-K\sqrt{L\log L})/(2\pi)$ imply $\beta\delta\,\Xi(\beta)^2\le c_1$.
\end{lemma}
\begin{proof}
(a) With $\beta_\delta:=(1-\varepsilon)L/(2\pi)\le L$,
\begin{align*}
\delta f(\beta)\le\delta f(\beta_\delta)&=\beta_\delta\exp\Bigl(-L+(1-\varepsilon)L+2C_\Xi\sqrt{(1+\beta_\delta)\log(2+\beta_\delta)}\Bigr)\\
&\le L\exp\Bigl(-\varepsilon L+2C_\Xi\sqrt{(1+L)\log(2+L)}\Bigr),
\end{align*}
which tends to $0$ as $L\to\infty$. (b) For $L\ge4$ we have $(1+L)\log(2+L)\le1.875\,L\log L$, hence
$2C_\Xi\sqrt{(1+L)\log(2+L)}\le2.74\,C_\Xi\sqrt{L\log L}$ and
\[
\delta f(\beta)\le L\exp\bigl(-(K-2.74\,C_\Xi)\sqrt{L\log L}\bigr)\le Le^{-2\sqrt{L\log L}},
\]
which tends to $0$.
\end{proof}

\subsection{General systems; proofs of Theorems~\ref{thm:main4} and~\ref{thm:phys}}\label{sec:general}

\emph{Box conditioning.} Let $\Sys$ be $(\beta,\theta)$-admissible, $x\in V\setminus P$, and $r\ge1$ an integer with
$\dist_\infty(x,P)\ge r$. Put $\Lambda^\circ:=\{z:\|z-x\|_\infty<r\}$, $R:=\{z:\|z-x\|_\infty=r\}$ and $\Lambda:=\Lambda^\circ\cup R$, a
translate of $Q(2r)$. Add the sites of $\Lambda\setminus V$ with independent $\omega_n$-distributed spins, and the missing
edges of $\Lambda$ with $b_e=0$ and $M_e=0$; the system stays admissible and the law of the original spins does not
change. By the Markov property of nearest-neighbour Gibbs measures (see~\cite[Chapter~2]{Georgii1988}),
given everything outside $\Lambda^\circ$ the spins in $\Lambda^\circ$ have the law $\langle\cdot\rangle_{\Lambda,\sigma|_R}$ of the
pinned-square system on $\Lambda$ with ring values $\sigma|_R$ (edges inside $R$ contribute constants). So for
bounded $F$ measurable with respect to the spins outside $\Lambda^\circ$ and a function $h$ of $\sigma_x$,
\begin{equation}\label{eq:boxcond}
\langle F\,h(\sigma_x)\rangle=\bigl\langle F\,\langle h(\sigma_x)\rangle_{\Lambda,\sigma|_R}\bigr\rangle.
\end{equation}

\emph{The case $n=4$.} If $\theta\,\Xi(\beta)^2\le c_1$, Lemma~\ref{lem:scale} and Proposition~\ref{prop:pinneddecay} with
$A=\{x\}$ and $d=r$ give, for every ring configuration $\varrho$ and every odd $O$,
\begin{equation}\label{eq:n4ring}
|\langle O(\sigma_x)\rangle_{\Lambda,\varrho}|\le2^{3+w}\|O\|_\infty e^{-r/\Xi(\beta)}.
\end{equation}
Put $A_*:=2^{3+w}$, which also covers the case $d\le2s(w+2)$ of Proposition~\ref{prop:pinneddecay}.

\emph{The case $n\ge5$, diagonal $M_e$.} Fix $i$ and a set $W_i\ni i$ of four coordinates, and write
$\sigma=(\rho u,v)$ with $u\in S^3\subset\R^{W_i}$, $v\in\R^{W_i^c}$ and $\rho=(1-|v|^2)^{1/2}$. By
Lemma~\ref{lem:gauss} with $k=4$, $\rho>0$ almost surely and under $\omega_n$ the direction $u$ is uniform on $S^3$
and independent of $v$. For diagonal $M_e$,
$\sigma_z^{\mathsf T}M_e\sigma_{z'}=\rho_z\rho_{z'}\sum_{k\in W_i}m_e^ku_z^ku_{z'}^k+\sum_{k\notin W_i}m_e^kv_z^kv_{z'}^k$, with no cross term.
Conditionally on $v$ in $\Lambda^\circ$ and on the ring values, $u$ is a pinned-square system with $n=4$, couplings
$M'_e=\rho_z\rho_{z'}\mathrm{diag}(m_e^k)_{k\in W_i}$, $b'_e=\rho_z\rho_{z'}b_e$, and pins $u_p$ (arbitrary where $\rho_p=0$, in which case
the couplings to $p$ vanish). It is $(\beta,\theta)$-admissible, since $|b'_e|\le\beta$ and
$\|M'_e-b'_eI\|\le\rho_z\rho_{z'}\theta\le\theta$. By~\eqref{eq:n4ring} applied to $u^i$,
$|\langle\sigma_x^i\rangle_{\Lambda,\varrho}|=|\langle\rho_x\langle u_x^i\mid v\rangle\rangle_{\Lambda,\varrho}|\le A_*e^{-r/\Xi(\beta)}$.

\begin{proof}[Proof of Theorem~\ref{thm:main4}]
(a) If $P=\emptyset$, the symmetry $\sigma\mapsto-\sigma$ gives $\langle O(\sigma_x)\rangle=0$; otherwise take $r=\dist_\infty(x,P)$
and $F=1$ in~\eqref{eq:boxcond} and use~\eqref{eq:n4ring}. (b) Take $r=\min\{\|x-y\|_\infty,\dist_\infty(x,P)\}$; then
$y\notin\Lambda^\circ$, and $F=O'(\sigma_y)$ is measurable outside $\Lambda^\circ$. (c) Let $d=d_T(x,y)$ and
$r=\min\{d,\lfloor(\min\{L_1,L_2\}-1)/2\rfloor\}$; if $r=0$ the bound is trivial since $A_*\ge1$. Otherwise
$2r+1\le\min\{L_1,L_2\}$, so the box $\Lambda$ around $x$ embeds in the torus without wrapping, $y\notin\Lambda^\circ$, and
the argument of (b) applies with the Markov property on the torus.
\end{proof}

\begin{proof}[Proof of Theorem~\ref{thm:phys}]
The system is $(\beta,\beta\delta)$-admissible, and by Lemma~\ref{lem:threshold}(a), with $\delta_0(\varepsilon)$ from there,
$\beta\delta\,\Xi(\beta)^2\le c_1$. For $n=4$ apply Theorem~\ref{thm:main4} with $O=\sigma^i$ and $O'=\sigma^j$. For $n\ge5$,
\eqref{eq:boxcond} and the bound on $\langle\sigma_x^i\rangle_{\Lambda,\varrho}$ above give (i) and (ii) in the same way. For
$P=\emptyset$ and diagonal $M_e$, flipping $\sigma^i$ at every site preserves the measure, so
$\langle\sigma_x^i\sigma_y^j\rangle=0$ for $i\ne j$. On a square torus $d_T(x,y)\le\lfloor L/2\rfloor\le\lfloor(L-1)/2\rfloor+1$, so
$r\ge d_T(x,y)-1$ and $e^{-r/\Xi}\le e\cdot e^{-d_T(x,y)/\Xi}$ since $\Xi\ge1$. For an infinite-volume Gibbs measure $\mu$
and every $r\ge1$, the DLR equations give $\mu(\cdot\mid\mathcal F_{\Z^2\setminus\Lambda^\circ})=\langle\cdot\rangle_{\Lambda,\sigma|_R}$ $\mu$-almost
surely; so $|\mu(\sigma_x^i)|\le A_*e^{-r/\Xi}$ for every $r$, hence $\mu(\sigma_x^i)=0$, and with $r=\|x-y\|_\infty$ the
argument of (ii) gives the two-point bound. The sharper condition on $\beta$ stated after
Theorem~\ref{thm:phys} is Lemma~\ref{lem:threshold}(b).
\end{proof}

For $n=4$ the same argument shows that every Gibbs measure of a $(\beta,\theta)$-admissible specification with
$\theta\,\Xi(\beta)^2\le c_1$ gives mean zero to every odd observable of one spin.

\subsection{What does not transfer}

\begin{remark}[only odd observables transfer]\label{rem:twosite}
For the isotropic model, \cite[\S3.10]{OpenAIO4} gives decay for every bounded observable whose average under
common left multiplication by $S^3$ vanishes. For anisotropic couplings this fails already on two sites.
Take the free two-site law on $S^3\times S^3$ proportional to $\exp(b\,u\cdot v+bt\,u^1v^1)$. Under the isotropic
law ($t=0$), $\E[v\mid u]=c(b)u$ with $c(b)=\langle u\cdot v\rangle=I_2(b)/I_1(b)>0$ for $b>0$, where $I_1,I_2$ are
modified Bessel functions of the first kind, and $\E(u^1)^4=\frac18$, $\E(u^1)^2=\frac14$. Hence
\[
\frac{d}{dt}\Big|_{t=0}\langle(u^1)^2\rangle=b\Cov\bigl((u^1)^2,u^1v^1\bigr)=b\Bigl(\frac c8-\frac14\cdot\frac c4\Bigr)=\frac{b\,c(b)}{16}\ne0.
\]
So the observable $(u^1)^2-\frac14$, whose average over common left rotations vanishes, has nonzero mean for
small $t\ne0$, with no pins at all; at $b=1$ the derivative is $0.01501\ldots$, which a direct quadrature
of the two-site integral confirms. Putting $b_e=0$ on every other edge embeds this example in pinned squares
of every size, within the class of Theorem~\ref{thm:main4}. Hence the restriction to odd observables cannot be
replaced by a vanishing $S^3$-average: even observables, such as bond energies or $(\sigma_x^1)^2$, are not
covered, and neither is uniqueness of the Gibbs measure beyond the vanishing of odd one-point functions. In
the proof, oddness enters only in Lemma~\ref{lem:odd}(c) and in the oddness step of
Proposition~\ref{prop:pinneddecay}, and evenness of the interaction only in Lemma~\ref{lem:pm1}(a). An odd
single-site term, such as a field $h\sum_x\sigma_x^1$ with $|h|\le\beta\delta$, satisfies
Lemma~\ref{lem:celltransfer} but not Lemma~\ref{lem:pm1}, and with it the conclusion of
Theorem~\ref{thm:main4}(a) fails.
\end{remark}

\begin{remark}[scope]\label{rem:scope}
Apart from the group structure of $S^3$ used for the coarse orientations, the value $n=4$ and the dimension
two enter only through (I1) and (I3), and the reduction from $n\ge5$ to
$n=4$ lowers $n$, so it cannot reach $n=3$. For $n=3$ the predicted correlation length
$e^{2\pi\beta}/\beta$ is much larger than $\Xi(\beta)$, so an argument at the scale $\Xi(\beta)$ cannot reach the
$O(3)$ crossover. For $n\ge5$ Theorem~\ref{thm:phys} gives the $O(4)$ threshold, below the prediction
$(n-2)|\log\delta|/(4\pi)$. Not covered are: truncated correlations of even observables; the regime above the
threshold, where no matching order statement is proved; non-diagonal anisotropy for $n\ge5$; single-site
anisotropy (odd single-site terms are excluded by Remark~\ref{rem:twosite}; even ones would add $O(\beta\delta s^2)$
per cell in Lemma~\ref{lem:celltransfer}, but this is not carried out here); and rectangular tori beyond the
scale $\min\{L_1,L_2\}/2$. The constants $c_1,A_*,C_\Xi,\delta_0(\varepsilon)$ are not numerical; they depend on the
fixed choices of~\cite{OpenAIO4}.
\end{remark}

{\small\raggedright

}

\end{document}